\documentclass[preprint2]{aastex}
\usepackage{epsfig}

\def\etal{et~al.\ }
\def\ltsima{$\; \buildrel < \over \sim \;$}
\def\simlt{\lower.5ex\hbox{\ltsima}}
\def\gtsima{$\; \buildrel > \over \sim \;$}
\def\simgt{\lower.5ex\hbox{\gtsima}}

\def\h1{\ion{H}{1}\ }

\def\h2{H$_2$}
\def\coh2{CO/H$_2$}

\journalid{}{}
\articleid{}{}
\shortauthors{Holley-Bockelmann \etal}
\shorttitle{Bars, Cusps and Cores}
\slugcomment{Draft - \today}
 
\begin{document}
 
\title{Bar-Induced Evolution of Dark Matter Cusps}
 
\author{Kelly Holley-Bockelmann, Martin D. Weinberg, and Neal
  Katz\altaffilmark{1}}

\altaffiltext{1}{ Department of Astronomy, University of
  Massachusetts, Amherst, MA 01003 kelly@shrike.astro.umass.edu,
  weinberg@astro.umass.edu, nsk@astro.umass.edu}

\begin{abstract}
  
  The evolution of a stellar bar transforms not only the galactic
  disk, but also the host dark matter halo. We present high
  resolution, fully self-consistent N-body simulations that clearly
  demonstrate that dark matter halo central density cusps flatten as
  the bar torques the halo. This effect is independent of the bar
  formation mode and occurs even for rather short bars.  The halo and
  bar evolution is mediated by resonant interactions between orbits in
  the halo and the bar pattern speed, as predicted by linear
  Hamiltonian perturbation theory.  The bar lengthens and slows as it
  loses angular momentum, a process that occurs even in rather warm
  disks.  We demonstrate that the bar and halo response can be
  critically underestimated for experiments that are unable to resolve
  the relevant resonant dynamics; this occurs when the phase space in
  the resonant region is under sampled or plagued by noise.

\end{abstract}
 

\section{Introduction}

It is widely accepted that a galactic bar will trigger a rearrangement of
the stellar and gaseous disk, but the bar's effect on the dark halo is
more controversial.  Linear Hamiltonian perturbation theory suggests
that the transfer of angular momentum drives post-formation galaxy
evolution, and is mediated by orbits that are in
resonance with quasi-periodic perturbers.  In the case of a barred galaxy, 
substantial amounts of
angular momentum are transferred from the bar to the halo via resonant
interactions between the bar pattern speed and the orbits of dark
matter particles in the inner halo (Lynden-Bell \& Kalnajs 1972,
Tremaine \& Weinberg 1984, Athanassoula 2000). The bar is a huge,
organized source of angular momentum, and the transfer of this angular
momentum from the bar to the disk--halo system causes the galaxy to
evolve. The coupling between the bar and halo, and the subsequent
evolution of both components, in particular the central dark matter
density profile, has been conclusively shown in idealized numerical
experiments (Weinberg \& Katz 2002, hereafter Paper I, Weinberg \&
Katz 2003, hereafter Paper II).

Unfortunately, attempts to study bar-halo interactions using fully
self-consistent N-body simulations have lead to wildly different
conclusions. For example, one robust prediction of this mechanism is
that the bar pattern speed slows as it loses angular momentum to the
halo, if the bar moment of inertia remains constant (Tremaine \&
Weinberg 1984, Weinberg 1985).  While the bar does appear to slow in
most N-body simulations (Athanassoula 2000, Hernquist \& Weinberg
1992, Sellwood 2002, Debattista \& Sellwood 2000), in others it does
not (Valenzuela \& Klypin 2002, hereafter VK).  The discrepancy could
be caused by differences in the numerical techniques, some of which
might be incapable of following the important physical processes, or
to differences in the disk and halo models.  These could affect either
the transfer of angular momentum from the bar to the halo, the
evolution of the bar moment of inertia, or both.

Even for simulations that produce a slowing bar, the effect on the
dark matter halo remains a point of contention. We have argued that
resonant coupling between the bar and halo can flatten central dark
matter density cusps (Hernquist \& Weinberg 1992, Paper I, Paper II).
The inner halo orbits gain enough angular momentum to move them to larger
time-averaged radii, which slowly removes the central density cusp.
This causes the center to lose gravitational support and accelerates
the flattening.  However, these results rely on idealized N-body
simulations, and to date there has been only one fully
self-consistent simulation that has explicitly shown the decrease in
the halo central density due to this process. Such evolution can be
seen in Figures 8 and 12 of Athanassoula (2003) although not mentioned
in the text, but other published simulations show that the central
density either remains unchanged (VK) or increases (Sellwood
2002). This has led many dynamicists to incorrectly conclude that
resonant interactions are unimportant in realistic galaxy evolution,
even though resonant dynamics has been shown to be crucial for bar
formation (Athanassoula 2002).

Clearly, one needs to verify the predictions of linear Hamiltonian
perturbation theory and those of idealized N-body simulations using
realistic, high-resolution, fully self-consistent N-body simulations.
This paper is one in a series designed to study bar-induced halo
evolution using realistic initial conditions and a state-of-the-art
N-body code that minimizes small scale noise and fully resolves the
galaxy disk.  Paper I (Weinberg \& Katz 2002) applied the concept of
bar-halo interactions to the evolution of the halo density profile
using idealized N-body simulations.  Paper II (Weinberg \& Katz 2003)
deals with common misconceptions about resonant dynamics and conducts
experiments to show that the predictions of linear Hamiltonian
perturbation theory pertain to idealized rigid bar and halo
simulations. In this work (Paper III), we attack the fully self-consistent
problem and demonstrate that resonant dynamics still applies to this
regime. We show that resonant interactions between the bar pattern
speed and orbits in the halo are critical to the evolution of a
fully self-consistent N-body experiment.

Our fiducial simulation has a bar length of about one disk scale length.
This bar length is comparable to those of recent bar-formation
simulations (VK, Sellwood 2002, Athanassoula 2003).  Using our
low noise N-body approach, we hope to better understand the large
differences among published self-consistent N-body simulations.
While it is difficult to definitively determine the cause of the
differences without direct access to either their initial conditions or
numerical codes, we argue that where
differences exist, they are attributable to subtle numerical artifacts
that destroy the important resonant dynamical processes.
These processes are well described by linear
perturbation theory, but the effect can be underestimated
either by compromising the resonant potential
through numerical noise, by having insufficient phase space coverage
near the resonance, or by artificially scattering the resonant orbit
reservoir.  We address the problems inherent to using any N-body
technique to track resonant dynamics in general, and discuss the
probable shortcomings of a few recent self-consistent simulations in
the discussion section.

Correctly following the physics on the `microscopic' level of an
individual orbit response can have effects on a galactic scale with
cosmological implications.  Cold dark matter (CDM) halos in every mass
regime are thought to form with a characteristic density profile,
expressed as $\rho(r) \propto r^{-\gamma} (1 + r/r_s)^{\gamma-3}$
(Navarro et al 1997, hereafter NFW).  While disagreements remain on
the precise value of the central slope, $\gamma$, which range between
$-1$ and $-1.5$ (Moore et al 1998, Power et al 2002), there is a
consensus that primordial dark matter halos are universally cuspy.
This prediction is testable using the rotation curves of galaxies.  By
determining the radial portion of the gravitational potential,
rotation curves when combined with light profiles can constrain both
luminous and dark matter density profiles. Though this
technique suffers from degeneracies in interpretation, many real dark
matter halos appear to be much less cuspy than those predicted by
standard CDM (de Blok et. al. 2001, McGaugh 2000, Swaters et. al.
2002), and some are even consistent with a flat central profile, or
density 'core'.

If the resonant dynamic processes described above occur in real
galaxies and if bars are a ubiquitous phase of early galaxy evolution,
the cusp--core controversy could be reconciled through subsequent
bar-halo interactions.  In this scenario, primordial disks form in a
cuspy dark matter halo. This proto-galactic disk is dynamically cold,
making it very susceptible to bar formation.  Such young galaxies are
subject to repeated satellite encounters (T\'oth \& Ostriker 1992,
Steinmetz \& Navarro 2002), and the first substantial one will likely
excite a large bar in the disk (Binney \& Tremaine 1987, Walker, Mihos
\& Hernquist 1996).  The length of the bar will depend on the mass and
distance of the satellite, i.e. the torque applied by an external
quadrupole. The typical bar induced by this process will be much
larger than those formed through internal disk instabilities, perhaps
even encompassing the entire disk. The number of halo orbits
commensurate with the bar pattern speed increases as the bar size
increases; for a massive primordial bar, the reservoir of resonant
halo orbits stretches from deep inside the halo cusp to well outside
the disk, allowing a broad range of radii to accept angular momentum
via resonant exchange. In idealized calculations, the torque from a
primordial bar can destroy an NFW cusp out to as much as half the bar
radius (Paper I, Paper II).

In this paper we emphasize moderate strength, scale-length-sized bars,
so our results set a $lower$ $limit$ to the halo evolution that would
be induced by tidal encounters in proto-disks.  Present-day galaxies
have smaller, weaker bars than those studied here and hence the
present-day effects could be smaller.  However, Sellwood (2002) argues
that both at high and low red shift, bar sizes are restricted to the
rising part of the rotation curve; this limits bar sizes to be less
than a disk scale length, thus invalidating the primordial scenario outlined
above.  We explicitly 
demonstrate that this is not true if the bars are triggered
externally by forming a long lived bar over four disk scale lengths in
size.  In addition, Athanassoula (2003) also forms bars of this size
without having to resort to external triggers. Furthermore, Jogee et
al (2002) show that the lengths of even local bars are severely
underestimated, and can extend to beyond the disk scale length.
Therefore, the scenario we outlined above remains a valid one.

We find that a bar with sizes of a disk scale length $can$ remove dark
matter cusps out to nearly 1/3 of the initial disk scale length using
simulations with 5 million equal mass dark matter halo particles within
the virial radius.  The orbits in
the central regions of the halo gain enough angular momentum to remove
the cusp.  We demonstrate the robustness of this result by obtaining
the same result with a simulation that uses 10 million particles.
Furthermore, in agreement with the predictions of linear Hamiltonian
perturbation theory, both these fully self-consistent simulations show
that low-order resonances are responsible for the transfer of angular
momentum from the bar to the halo.  We explicitly show that the
angular momentum is deposited at discrete resonances in phase space,
which counters some claims that resonant dynamics is only important in
idealized situations.  For our self-consistent field (SCF) code, we find
that simulations using only 1 million particles have insufficient
phase space resolution (Paper II) and too much discreteness noise to
resolve the important resonances, and hence cannot follow
the relevant bar-halo physics correctly. The required number of 
particles may be even higher for
other numerical N-body techniques, which have more small scale noise.
Interactions with satellites and 
more distant
group members can also drive resonant angular momentum
exchange, but are more complicated since they involve more disparate
time scales.  It is likely that such simulations would require many more
particles than the simple bar perturbation case studied here. 

We organize this paper as follows. \S1 describes the models, \S2 reviews
the important physical processes, \S3
outlines our N-body technique, how we realize our initial conditions
and the specific set of experiments we perform.  We present our 
results in \S4, where we describe the bulk changes to the system, discuss the
angular momentum exchange, identify the resonant interactions
responsible for the halo evolution, and discuss the effect on the both
the halo and disk density profiles. \S5 investigates whether our
results are effected by numerical artifacts: we
confirm that the system is stable against centering instabilities,
explore the effect of the bar formation mechanism, and discuss the
effect of particle number.  We 
present the implications of our results
in \S6, where we detail to what extent our final evolved halos
resemble observed galaxies, and discuss the reasons for any disagreements
with past work. \S7 summarizes and outlines future work.

\section{Physical motivation}

Here we outline the physical processes and numerical considerations
important to the study of bar interactions.  A much more detailed
discussion can be found in \S2 of Paper II.  In a near-equilibrium
galaxy, a global and lasting change in any conserved quantity can only
occur at resonances, integer commensurabilities between orbital
frequencies and a perturbation frequency.  A bar is a natural rotating
disturbance that resonantly exchanges angular momentum with both halo
and disk orbits.  To an observer sitting on the bar, these
commensurate orbits describe closed, non-axisymmetric figures.
Because the figures are non-axisymmetric, they will be torqued by the
bar.  Orbits that are slowly precessing in the bar frame, on either
side of the commensurability, can also receive a torque for a
finite time.  But no matter how slowly an orbit processes, if one
waits long enough in a fixed potential with a constant bar pattern
speed, the slow precession of the orbit eventually will describe an
axisymmetric figure and would produce zero net torque.  However, this
time will be longer than times of interest in galaxy evolution, so in
a practical sense, many orbits will be torqued.  In addition, the bar
pattern speed and gravitational potential are slowly evolving and,
therefore, a precessing orbit may become closed and begin precessing
in the opposite direction in response to this slow change.  Such an
orbit has {\em passed through the resonance.}  At the point that the
orbit is closed, adiabatic invariance is broken and the orbit is
sensitive to the gravitational attraction of the bar through the
potential associated with the resonance.  Only those orbits that pass
through the resonance irreversibly change their actions.  The galaxy
equilibrium as a whole then slowly evolves owing to these changes, and
this brings fresh orbits through the resonance to continue the
evolution.

The fractional change in the angular momentum of an individual orbit
is antisymmetric about location of the resonance.  Unless there is a
gradient in phase space, as the resonance sweeps through the system
orbits will pass through the resonance in both directions, and the sum
of these individual changes over the ensemble of orbits containing the
resonance will cancel (e.g. Weinberg 2001).  A phase space gradient
causes an incomplete cancellation that gives rise to a net torque.
The total angular momentum gains and losses at a given time for
individual orbits are first order in the bar amplitude, while the net
change in the halo's angular momentum is second order and results from
the near cancellation of this first order effect.  The specific net
change, therefore, is much smaller than the average change for an
individual orbit.

A minimum requirement to accurately model resonant dynamics is that
the resonant potential is well-populated by particle orbits. 
Since the change in any individual orbit as it passes through the resonance
depends on the phase of the orbit, an accurate result
relies on a dense sampling in phase.  For a given resonant potential,
there is a critical number of particles, $f_{\rm crit}$, required to
resolve the region near the resonance such that the response is
represented by contributions from orbits at many phases. As explained in the 
previous paragraph, the change in
the actions of any one orbit is first order in the bar amplitude, but
these changes must cancel over the ensemble of orbits, leaving only
the smaller second-order contributions that are responsible for
long-term halo evolution.  If phase space is incompletely sampled, the
first-order changes will {\em not} cancel, but will produce random
fluctuations and the real resonance-driven evolution will not occur.

Even when the resonance potential is well-resolved by particle orbits,
it may still be rendered ineffective if potential fluctuations caused
by the finite-number of particles swamp the resonant potential.  It is
possible to determine the power in the Poisson noise at any scale by
Fourier analysis, or more generally from any orthonormal basis
expansion that satisfies the Poisson equation in an axisymmetric
coordinate system (Weinberg 1998). To resolve the resonance potential
in the basis expansion, we require that the power in the coefficients
of the halo response be greater than the power in the noise.  Since a
basis expansion is at the heart of our potential solver, we can use
this signal-to-noise criterion to determine the critical number of
particles needed to resolve the resonance potential for any expansion
term used in our N-body simulation (see Paper II).

Noise on interparticle scales scatters orbits from their original
trajectories and leads to a diffusion time scale that is unphysically
short.  If the diffusion time is shorter than the time that
potentially resonant orbits would need to pass through a resonance
driven by the slow evolution of the galaxy, the resonance will cease
to exist.  Without these stable, resonant orbits, the exchange of
angular momentum through this mechanism cannot take place.  This
diffusion on interparticle scales, i.e. {\em two-body relaxation}, is
astronomically negligible for our bar--halo simulations, but can be a
significant problem for other studies due to the numerical noise
present in many N-body potential solvers.  Since the phase-space width
of the resonance is proportional to the strength of the bar
perturbation, the time scale for drifting across the resonance can be
a small fraction of the two-body relaxation time scale.  However, direct
summation, tree, and grid-based codes are all particularly prone to
generating small-scale noise and are thus quite susceptible to rapid orbit
diffusion.  The only remedy is to increase the particle number while
keeping the resolution fixed, thereby pushing the simulation toward
the collisionless limit.  Expansion codes still suffer from relaxation
that will diffuse orbits in principle, but the {\em small-scale} noise
is removed by truncating the expansion.

\section{Numerical Techniques}

\subsection{N-body code}

The disk and dark matter halo are evolved using a 3-dimensional
self-consistent field (SCF) code (Weinberg 1999). In most N-body
methods, it is either the gravitational softening introduced to
decrease two-body scattering, or the grid cell size that determines
the spatial resolution.  In such codes, the number of spatial
resolution elements within the simulation volume determines the
effective number of degrees of freedom, typically a very large number.
SCF codes (Earn \& Sellwood 1995, Clutton-Brock 1972, 1973, Kalnajs
1976, Polachenko \& Shukmann 1981, Friedman \& Polyachenko 1984,
Hernquist \& Ostriker 1992, Hernquist, Sigurdsson, \& Bryan 1995,
Brown \& Papaloizou 1998, Earn 1996, Allen, Palmer, \& Papaloizou
1990, Saha 1993) limit the number of degrees of freedom to decrease
the small-scale noise, making this class of
code ideal for the simulation of the long term evolution caused by
resonant dynamics.

Our potential solver exploits properties of the Sturm-Liouville
equation to generate a numerical bi-orthogonal basis set whose lowest
order basis function matches the equilibrium model (Weinberg 1999).
Many important physical systems in quantum and classical dynamics
reduce to the Sturm-Liouville (SL) form,
\begin{equation}
{{d}\over {dx}}{ \Big[ p(x) {{d\Phi(x)} \over {dx}} \Big] -q(x) \Phi(x)} = 
{\lambda \omega(x)\Phi(x)},
\end{equation}
where $\lambda$ is a constant and $\omega(x)$ is a known function,
called either the density or weighting function.  If $\Phi(x)$ and
$\omega(x)$ are positive in the interval $a<x<b$, then the SL equation
is satisfied only for a discrete set of eigenvalues, $\lambda_n$,
with corresponding eigenfunctions $\phi_n(x)$. The
eigenfunctions form a complete basis set (Courant \& Hilbert 1953) and
can be chosen to be orthogonal with the following additional
properties: 1) the eigenvalues $\lambda_n$ are countably infinite and can
be ordered: $\lambda_n < \lambda_{n+1}$; 2) there is a smallest
non-negative eigenvalue, $\lambda_1 > 0$, but there is no greatest
eigenvalue; and 3) the eigenfunctions, $\phi_n(x)$, possess nodes between
$a$ and $b$, and the number of nodes increases with increasing $n$,
e.g. the eigenfunction $\phi_1(x)$ has no nodes, $\phi_2(x)$ has one
node, etc.

In the special case of Poisson's equation, we use the eigenfunctions
to construct biorthogonal density and potential pairs, $d_k$ and $u_j$,
given by:
\begin{equation}
{1/{4 \pi G} \int dr r^2 d_k^*(r) u_j(r)} = {\delta_{jk}}.
\end{equation}
The lowest order potential-density pair ($n=1$, $l=m=0$) represents the
equilibrium profile, and the higher order terms represent deviations
about this profile.

For the equilibrium profile, we use a numerically relaxed and truncated
form of the NFW profile as we describe below.  We retain halo basis
terms up to $n_{\rm max}=10$ and $l_{\rm max}=4$. For the disk, we
derive a cylindrical basis from a spherical SL basis, using the
empirical orthogonal function method described by Weinberg (1996,
1999). This method constructs a new linear combination of the original
basis to closely match the disk profile.  We derive the spherical disk
SL basis from the deprojected disk profile using a large number of
terms ($n_{\rm max}=10, l_{\rm max}=36$) to accurately resolve the
thinness of the disk. Weinberg (1999) used a
cylindrical solution of the SL equation to construct the empirical
basis, a solution
whose boundary conditions are inconsistent with the spherical
halo basis; this motivated our use of a spherical SL basis for the disk 
as well.
We retained ten basis functions per azimuthal
harmonic order, $m$, from this empirical basis set. The expansion
parameters for each simulation (described below) are shown in
Table \ref{tab:simparams}.  Particles are
advanced using a leapfrog integrator, with a time step $h=0.0002$,
which is $0.008$ of the smallest oscillatory period, the orbital
frequency in the very central region, or
$0.009$ $\rm{t_{\rm dyn}}$ at a disk scale length.

\begin{deluxetable}{rrrrrrrrrr}
\tablecolumns{10} 
\tablewidth{0pc} 
\tablecaption{Simulation parameters} 
\tablehead{ 
\colhead{Run} 
& \colhead{Component} 
& \colhead{${\rm n_{\rm max}}$} 
& \colhead{${\rm l}$}   
& \colhead{${\rm m_{\rm max}}$}   
& \colhead{${\rm Mass}$}    
& \colhead{${\rm R_{\rm d}}$} 
& \colhead{${\rm z_{0}}$}    
& \colhead{${\rm R_{h}}$}
& \colhead{c}

}
\startdata 
F,I,C,B & Disk\tablenotemark{a} & 10 & 0-36      & 4 &  0.06 & 0.01 & 0.001 &  
& \\
L       & Disk\tablenotemark{a} & 10 & 0-36      & 4 &  0.06 & 0.01 & 0.001 &  
& \\
F,I,C,B & Halo & 10 & 0,1,2,3,4 & 4 & 1 & &  & 1& 15 \\ 
L       & Halo & 10 & 0,2,3,4 & 4 & 1 & &  & 1& 15 \\ 
\enddata 
\tablenotetext{a}{Only the first $N_{\rm order}=12$ basis terms were included 
in the expansion. See Weinberg (1996) for details.}

\label{tab:simparams}
\end{deluxetable} 

\subsection{Generating initial conditions}

Our galaxy models have two components: a dark matter halo and an
initially bar-less, bulge-less exponential disk.  The axisymmetric
disk density profile is:
\begin{equation}
{\rho_d(R,z)} = {{M_d \over 8 \pi z_0 {R_d}^2} e^{-R/R_d} 
\hbox{sech}^{2}(z/z_0)},
\end{equation}
where $M_d$ is the disk mass, $R_d$ is the disk scale length, and
$z_0$ is the scale height. We truncate the disk at $20 R_d$ and at $10 z_0$.
Since the excluded mass is so small, we do not adjust the total
mass to account for this truncation.

We adopt an axisymmetric velocity dispersion in the disk plane
($\sigma_r = \sigma_\phi$) and set the radial velocity dispersion so
that the Toomre stability parameter, Q, is constant at all disk radii.
Hence
\begin{equation}
{\sigma_r}^2(R)= {Q} {{3.36 \Sigma(R)}\over { \kappa(R)}},
\end{equation}
where G=1, $\Sigma(R)$ is the surface density, and to produce good
equilibrium models, $\kappa$, the epicyclic frequency
\begin{equation} 
{{\kappa}^2(R)}= {R {{d\Omega^2}\over {dR}} + 4 \Omega^2},
\end{equation}
where $\Omega$ is the circular frequency, is derived from the actual
particle distribution.  We determine the vertical velocity dispersion,
$\sigma_z$, by solving Jeans' equations in cylindrical coordinates
assuming a steady-state disk:
\begin{equation} 
\sigma_z^2(R)= {1\over\rho_d(R,z)}\int_z^\infty{\rho_d(R,z){\partial\Phi_{tot}
\over \partial z} dz},
\end{equation}
where $\Phi_{tot}$ is the combined disk and halo potential.
The mean radial and vertical velocities are zero, and the mean
azimuthal velocity, $\bar V_\phi$, 
is determined from the asymmetric drift equation:
\begin{equation} 
  {\bar V_\phi^2}(R) = V_{\rm circ}^2(R) + {R\over \rho(R)} {d\over
      dR}\left[\rho(R) \sigma^2_r(R)\right],
\end{equation}
where $V_{\rm circ}$ is the circular velocity derived from the
combined halo-disk potential.  We realize the velocities by
approximating their distributions as Gaussians with the means and
dispersions given above.

The dark matter halo density distribution is a truncated NFW profile.
We use units where both the halo mass, $M_{\rm vir}$, and the virial
radius, $R_{\rm vir}$, are one.  Using these units, the density profile is:
\begin{equation}
{\rho(r)}= {{c^2 g(c)} \over {4 \pi r (1 + c r)^2}},
\end{equation}
where
\begin{equation}
{g(c)}= {{1} \over {\ln(1+c) - c/(1+c)}}.
\end{equation}
and $c$ is the concentration parameter, 
\begin{equation}
c=R_{\rm vir}/r_s,
\end{equation}
where $r_s$ is the scale radius.

Since the NFW mass profile is logarithmically divergent at large
radii, we truncate it at $R_{\rm vir}$. However, an abrupt truncation
of the density profile at the virial radius leads to a poor
equilibrium, given our assumed isotropic velocity distribution. To
generate a better equilibrium, we compute a phase-space distribution
function with an isotropic velocity distribution by Eddington
inversion (see Binney \& Tremaine 1987) for the truncated
halo and spherically averaged disk system combined.  We generate a 
new halo density profile by integrating this phase-space distribution 
function over
velocity.  If an equilibrium model with an isotropic velocity
distribution existed for the density profile, the new density profile
would be identical.  We repeat this procedure until the derived
density profile does not change, which only takes a few iterations.
The final halo density profile begins to noticeably deviate from a
pure NFW profile at approximately $90 \% R_{\rm vir}$.  We adjust the
mass so that the total halo mass after this truncation procedure does
not change.  We realize the halo particle distribution and velocities
simultaneously by Monte Carlo rejection using the phase-space
distribution function.

Because we are investigating the long-term evolution of a galaxy, we
take particular care to realize a stable equilibrium.  Unfortunately,
the stability of a cuspy profile is difficult to maintain with a
finite number of particles, since the inner cusp will always be poorly
resolved within some radius.  This resulting lack of gravitational
support causes the inner density profile to flatten within the poorly
resolved region.  To mitigate this problem, we generate our initial
conditions in a three-step process.  First, we populate the disk and
halo phase space as described above.  Then, we fix the disk potential
and evolve the composite system for several dynamical times.
This allows the inner halo to achieve a self-consistent equilibrium in
the presence of the disk.  The density profile turns over inside the
empirically-determined `resolution limit' of system, although it
retains its NFW profile outside this radius. Typically, this
resolution limit encloses about 20 particles out of our 5 million
particle realization.  Finally, we re-realize our initial conditions
using this new self-consistent halo density profile. This process results in
initial conditions that are remarkably stable in isolation; the Virial
relation $2T/VC$, where $VC$ is the Virial of Clausius, deviates from
unity by less than $0.1\%$ over $100$ dynamical times at the disk
scale length.  For an 11 million particle simulation, the
turn-over in the dark matter density profile occurs at approximately $1.5$ x
$10^{-4} R_{\rm vir}$, and occurs at $2.0\times 10^{-4} R_{\rm vir}$ and
$2.5 \times 10^{-4} R_{\rm vir}$ when using 5.5 and 1.1 million particles,
respectively.

\subsection{Bar formation}

Bars can be triggered through global disk instabilities (Toomre 1964),
or through
secular growth (e.g.  Polyachenko 1995).  In N-body simulations of
realistically hot disks, both mechanisms can be overwhelmed by density
perturbations that are caused by Poisson noise due to a finite-particle
realization of the potential (Ostriker \& Peebles 1973, Fall \&
Efstathiou 1980, Sellwood 1996).  This may or may not reflect bar
formation in nature, but does result in a large range in bar formation
times.  To control the onset time of bar formation, we trigger the bar by
applying an
external quadrupole potential.  We adopt the quadrupole profile for a
homogeneous ellipsoid with axis lengths $a: b: c = 0.02: 0.01: 0.001$
in model units. During the time the quadrupole is applied, it rotates
at a fixed pattern speed with corotation at the semi-major axis of
$a=0.02$. The perturbation has the form $Y_{22}(\theta, \phi)
\Phi_{2}(r)$, with
\begin{equation}
{\Phi_2(r)} \propto  
- A(t) { r^2  \over {[1.0 + (r/b_5)^{5/\alpha}]^\alpha}}.
\label{eq:quadpot}
\end{equation}
The quantity $b_5$ describes the characteristic radius of the
quadrupole; for $r\ll b_5$ ($r\gg b_5$), the quadrupole matches the
inner (outer) solution of the Laplace equation.  The exponent $\alpha$
determines the steepness of the central quadrupole potential; the
sharpness of the turn over between the inner and outer Laplace
solutions increases as $\alpha$ increases.  The quadrupole fit to a
homogeneous ellipsoid gives $\alpha\approx5$.  We adopt a time-varying
amplitude of
\begin{equation} 
 $$\displaylines{ A(t)= 0.25 A_0 \left[ 1.0+{{\rm erf}\left({{t-t_{\rm start}}\over
          {t_{\rm grow}}}\right)} \right]\times\cr 
\hskip90pt \left[ 1.0 - {{\rm erf}
      \left({{t-t_{\rm end}} \over {t_{\rm grow}}}\right)} \right].}$$\end{equation}
We choose $A_0$ and $a$ (which determines $b_5$) to match the
quadrupole strength and length of a purely noise-driven bar in a 5.5
million particle simulation.  We choose $\alpha=1$, but the final bar
profile is not sensitive to this value (see \S5.2 ).  For
nearly all the externally triggered quadrupole runs, $A_0=0.4$, $a=0.02$,
$t_{\rm start}=0.05$, $t_{\rm end}=0.08$, and $t_{\rm grow}=0.03$.
We choose the growth and damping time to be the dynamical time at the
bar length,
maximizing the orbit trapping efficiency of the external
perturbation. This quadrupole perturbation is only non-negligible
between at times $0.03-0.10$ and, hence, the bar evolves 
self-consistently for $t\gtrsim0.10$.

\subsection{Experiments}

\begin{figure}[t]
\epsfig{file=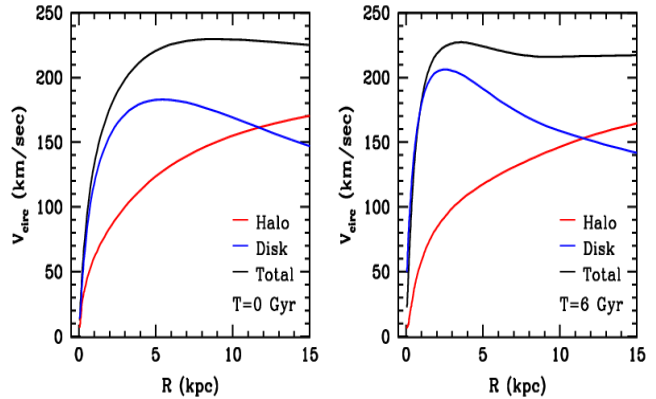, height=170pt, width=250pt}
\caption{Initial and final rotation curves for our fiducial
  simulation. The model is scaled to Milky Way units to facilitate
  comparison to our galaxy. The black line is the total rotation 
  curve, the red line
  represents the halo, and the blue line represents the disk. The
  inner halo expands while the disk contracts as discussed in \S 4. }
\label{fig:rotcurve}
\end{figure}

We adopt a halo concentration of $c=15$, consistent with CDM N-body
simulations that take into account angular momentum transfer between
the dark matter halo and baryons during galaxy formation (e.g. Jing
2000, Bullock et al 2001).  We choose our units as follows: the halo
mass is $1.0$, the virial radius $R_{\rm vir}$ is $1.0$, the disk
scale length $R_d$ is $0.01$, the scale height $z_0$ is $0.001$, and
the total disk mass $M_d$ is $0.067$. In these units, the halo scale
radius $r_s=0.067$. These parameters are summarized in Table
\ref{tab:simparams} and results in model galaxies with submaximal
disks as shown in Figure \ref{fig:rotcurve} (left).  To make it easier
to compare with observed galaxies, we also scale the simulations to a
typical dwarf galaxy and to the Milky Way.  For the dwarf galaxy, we
choose a circular velocity at the virial radius of 43 km/sec.  For the
Milky Way, we use the cosmological simulations of the local group by
Moore et al (1998) as a guide, making the circular velocity at the
virial radius $135$ km/sec.  All the resulting scalings are presented
in Table \ref{tab:conv}.  Throughout the text, whenever we quote a
system unit, we will follow it by the equivalent dwarf and Milky Way
scale in parentheses, i.e.  (dwarf unit, Milky Way unit).  For
example, in the 5.5 million particle halo, the finite particle induced
turn-over in the density profile occurs at $2.0 \times 10^{-4}$ (12
pc, 60 pc).

\begin{deluxetable}{rrrrr}
\tablecolumns{4} 
\tablewidth{0pc} 
\tablecaption{Model to Physical Unit Conversions} 
\tablehead{ 
\colhead{Galaxy} 
&\colhead{Length} 
& \colhead{Mass} 
& \colhead{Velocity} 
& \colhead{Time}
\cr \colhead {Type}
   }

\startdata 
Model & 1.0\phn & 1.0\phn\phn\phn\phn\phn & 1.0\phn\phn\phn & 1.0\phn\phn\\
Dwarf & 58 kpc & $2.5$ x $10^{10}$ ${\rm M_{\odot}}$ & 43 km/sec & 1.3 Gyr \\
${\rm Milky Way}\phn$ & 300 kpc  & $1.3$ x $10^{12}$ ${\rm M_{\odot}}$ & 135 
km/sec & 2.2 Gyr \\

\enddata
\label{tab:conv}
\end{deluxetable} 

We conducted 5 sets of simulations.  The fiducial runs (Set F) include
the external quadrupole trigger in a fully self-consistent halo and
disk model.  We also run fixed-disk control simulations (Set C) for a
consistency check on halo evolution. To ensure that the external
quadrupole is not inducing the halo to evolve more than that of a
bar formed
through disk instabilities, we run fully self-consistent simulations
that allow the bar to form on its own (Set I). Set L investigates 
a possible numerical artifact that could be
introduced by the $l=1$ portion of
the potential expansion, and Set B tests the assertion (Sellwood
2002) that disks cannot form lasting bars much longer than the disk
scale length.  Each simulation is run for at least one time unit (1.3
Gyr, 2.2 Gyr) and as many as three time units (3.9 Gyr, 6.6 Gyr).
Within each set, we
vary only the particle number. The minimum number of halo particles
used in any set is $N_{\rm halo}=10^6$, the maximum is $N_{\rm
halo}=10^7$, and the disk particle number is always chosen to be $
N_{\rm halo}/10$.  The subscript on the label for a particular run
refers to the number of halo particles in units of one million.  See
Table \ref{tab:barparams} for a synopsis of the experiments.

\begin{figure*}[t]
\epsfig{file=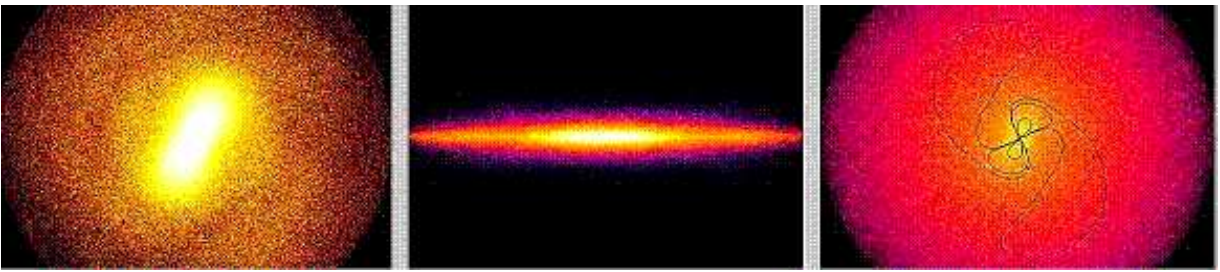,height=100 pt,width=460 pt}
\caption{Surface density maps for the $F_{10}$ run at $T=0.18$ (234 Myr,
396 Myr).
  The brightness corresponds to the logarithm of the density with
  white being the most dense and black being the least dense over 5
  orders of magnitude.  The horizontal scale for each panel is 10
  $R_d$ (5.8 kpc, 30 kpc).
  The left panel is the face on view of the stellar
  component and the center panel shows the disk edge on. The right
  panel is the face on view of the dark matter particles, with a
  superimposed contour plot of the $m=2$ component of the halo
  potential to accentuate the bar wake.}
\label{fig:tipsyden}
\end{figure*}

\section{The Fiducial Run}

A bar clearly forms in our fiducial simulation ($F_{5}$). Figure
\ref{fig:tipsyden} shows the face-on and edge-on view of the $F_{5}$
bar 0.18 time units (234 Myr, 396 Myr) after formation, as well as the wake
induced in the dark matter halo by the bar. 

The existence of a bar in such an initially cuspy system is a
relatively new result; the large central mass of a cuspy halo can
provide a barrier that prevents $X_1$ orbits from passing through the
center (Binney \& Tremaine 1987, Polyachenko \& Polyachenko 1996).  By
removing such a major bar orbit family, this spherical potential
barrier was thought to prevent the growth of a weak bar instability.
Recent work by Athanassoula (2002) suggests that, contrary to
hindering bar growth, cuspy dark matter halos encourage bar formation
via resonant interactions that transfer angular momentum from the disk
to the halo, consistent with our findings here.

In this section, we examine in detail the structural and kinematic
evolution of the disk-halo system in our fiducial simulation.
In addition, we explicitly demonstrate 
the importance of resonant
dynamics in this system through an analysis of halo orbits
in resonance with the bar pattern speed.

\subsection{Bulk characteristics}

\begin{figure}[t]
\plotone{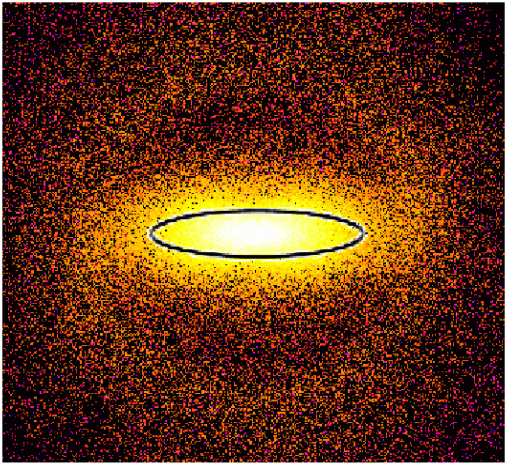}
\caption{The bar length estimated from the best fit ellipsoid  (black line)
to the quadrupole part of the potential is over plotted on the surface
density projection of the disk. Color scale is as described in Figure \ref{fig:tipsyden}. }
\label{fig:barlengthtip}
\end{figure}

Determining the bulk properties of the bar is a notoriously ambiguous
process.  For example, several methods have been proposed to define
bar length, ranging from the radius where there is an abrupt change in
the measured ellipticity of the disk, to the radius where the amplitude
of the $m=2$ to $m=0$ component of the disk drops below an empirically
tested threshold. We choose to adopt as our bar parameters the length,
mass, and ellipticity of a homogeneous ellipsoid fit to the $m=2$
component of the disk potential (see eqn. \ref{eq:quadpot}), fixing
the vertical height of the bar such that $c=b/10$.  (See Figure
\ref{fig:barlengthtip} for a comparison of the projected best-fit
ellipsoid and the projected bar surface density.)  The bar has an
initial length of 0.015 (0.87 kpc, 4.5 kpc), which is $1.5 R_d$ for
the initial disk scale length of 0.01 (0.58 kpc, 3.0 kpc).  The
initial mass of the bar is $0.2 M_d$ and the initial axis ratio is
$b:a=1:5$.  The initial and final bar parameters for all the
simulations are presented in Table \ref{tab:barparams}

\begin{deluxetable}{rrrrrrrrrrr}
\tablecolumns{10} 
\tablewidth{0pc} 
\tablecaption{Bar Parameters} 
\tablehead{ 
\colhead{Run} 
& \colhead{${\rm N_{\rm d}}$} 
& \colhead{${\rm N_{\rm h}}$} 
& \colhead{ ${\rm l_{\rm B,init}}$\tablenotemark{a,c}}
& \colhead{ ${\rm M_{\rm B,init}}$\tablenotemark{b,c}}
& \colhead{ ${\rm b/a_{\rm init}}$\tablenotemark{c}}
& \colhead{ ${\rm l_{\rm B,final}}$\tablenotemark{d}}
& \colhead{ ${\rm M_{\rm B,final}}$\tablenotemark{b}}
& \colhead{ ${\rm b/a_{\rm final}}$}
& \colhead{ Time}
&\colhead{Cusp}
\cr &&&&&&&&&&\colhead{      Disrupted? }
   }

\startdata 
${\rm F_1}$ & $ 1$ x $10^5$ 
& $1$ x $10^6$ & 0.018 & 0.018 & 1/4 & 0.024 & 0.03 & 1/5 & 2.38 & no\\ 
${\rm F_5}$ & $ 5$ x $10^5$ & $5$ x $10^6$ & 0.02 & 0.02 & 1/5 & 0.026 & 
0.025 & 1/6 & 3.0 & yes\\
${\rm F_{10}}$ & $ 1$ x $10^6$ & $1$ x $10^7$ & 0.02 & 0.021 & 1/4 & 0.025 & 
0.024 & 1/6 & 1.0 & yes\\
${\rm I_{5}}$ & $ 5$ x $10^5$ & $5$ x $10^6$ & 0.021 & 0.02 & 1/4 & 0.023 & 
0.022 & 1/5 & 1.5 & yes \\
${\rm L_{5}}$ & $ 5$ x $10^5$ & $5$ x $10^6$ & 0.017 & 0.018 & 1/5 & 0.023 & 
0.026 & 1/5 & 1.8 & yes \\
${\rm B_{5}}$ & $ 5$ x $10^5$ & $5$ x $10^6$ & 0.035 & 0.03 & 1/6 & 0.04 & 
0.035 & 1/6 & 2.4 & ?\\

\enddata
\tablenotetext{a}{The initial $R_d=0.01$}
\tablenotetext{b}{$M_d = 0.067$}
\tablenotetext{c}{Here, the inital values are measured at $t=0.2$}
\tablenotetext{d}{The final $R_d=0.004$}
\label{tab:barparams}
\end{deluxetable} 

\begin{figure*}[t]
\epsfig{file=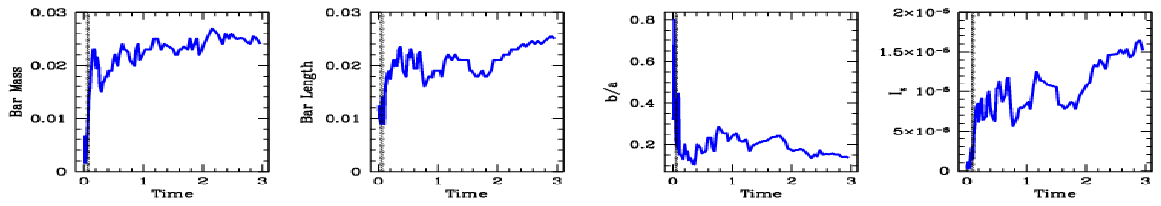,height=150 pt,width=500 pt}
\caption{The bar mass (far left panel), bar length (left center),
minor to major axis ratio (right center), and the z-component of the 
moment of inertia (far right) as a function of time for
the fiducial run ($F_5$). The vertical line is the time when the
triggering quadrupole stops.}
\label{fig:bulkbar}
\end{figure*}

Figure \ref{fig:bulkbar} shows the time evolution of these quantities.
Both the bar length and the mass grow by $60 \%$ over the course of
the experiment, though the bar figure remains relatively stable. At
first the axis ratio becomes somewhat rounder, going from 1/6 to 1/4
and then elongates once again to 1/6.  We also show the time evolution
of the z-component of the moment of inertia, which nearly doubles.  At
later times, the moment of inertia of the bar increases as more
particles from larger radii join the bar pattern, i.e. as the bar
lengthens.  The bar pattern also rapidly slows; over the course of the
simulation, the bar slows to about $30 \%$ of its original rotation
speed. A common way to express the rotational speed of the bar is by
the quantity $D_L/a_B$, where $D_L$ is the radius of corotation and
$a_B$ is a measure of the length of the bar's semi-major
axis. Initially, $D_L/a_B$ is nearly 1, consistent with the pattern
speeds in barred systems such as NGC 1365.  However, by the end of the
simulation, $D_L/a_B$ rises to 2.2, far slower than any observed bars
(Debattista \& Williams 2000, Merrifield \& Kuijken 1995, van Albada
\& Sanders 1982). Our strongly slowing bars are consistent with
Debattista \& Sellwood (2000) who found that bars slow significantly
in a cuspy dark matter halo.  If the moment of inertia were constant,
the drastic slowing of the bar would imply that it lost $70 \%$ of its
original angular momentum.  The doubling of the moment of inertia
lowers this angular momentum loss to about $40 \%$ of its initial
value.

\begin{figure}[t]
  \plotone{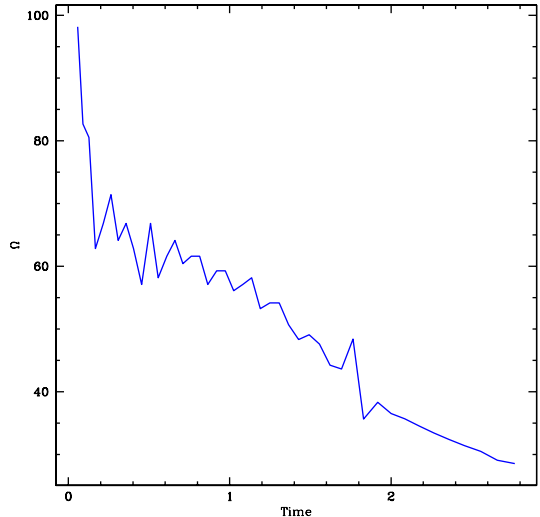}
\caption{The pattern speed of the bar as a function of time for simulation
$F_5$.}
\label{fig:barfreq}
\end{figure}

\subsection{Angular momentum deposition and the slowing of the bar}

The evolution of any collisionless system is governed by the transfer
of angular momentum between global modes, or patterns, and individual
orbits.  In the case of a barred galaxy, angular momentum exchange
between the bar and the halo or outer disk facilitates
the bar's formation (Athanassoula 2002), and causes the bar to slow
its rotation (Tremaine \& Weinberg 1984, Weinberg 1985, Debattista \&
Sellwood 2000, Athanassoula 2003). However, previous self-consistent
simulations have disagreed on the role of angular momentum transfer,
the relative importance of the disk-to-halo angular momentum exchange,
and the magnitude of the resultant bar slowing.

\begin{figure}[t]
\plotone{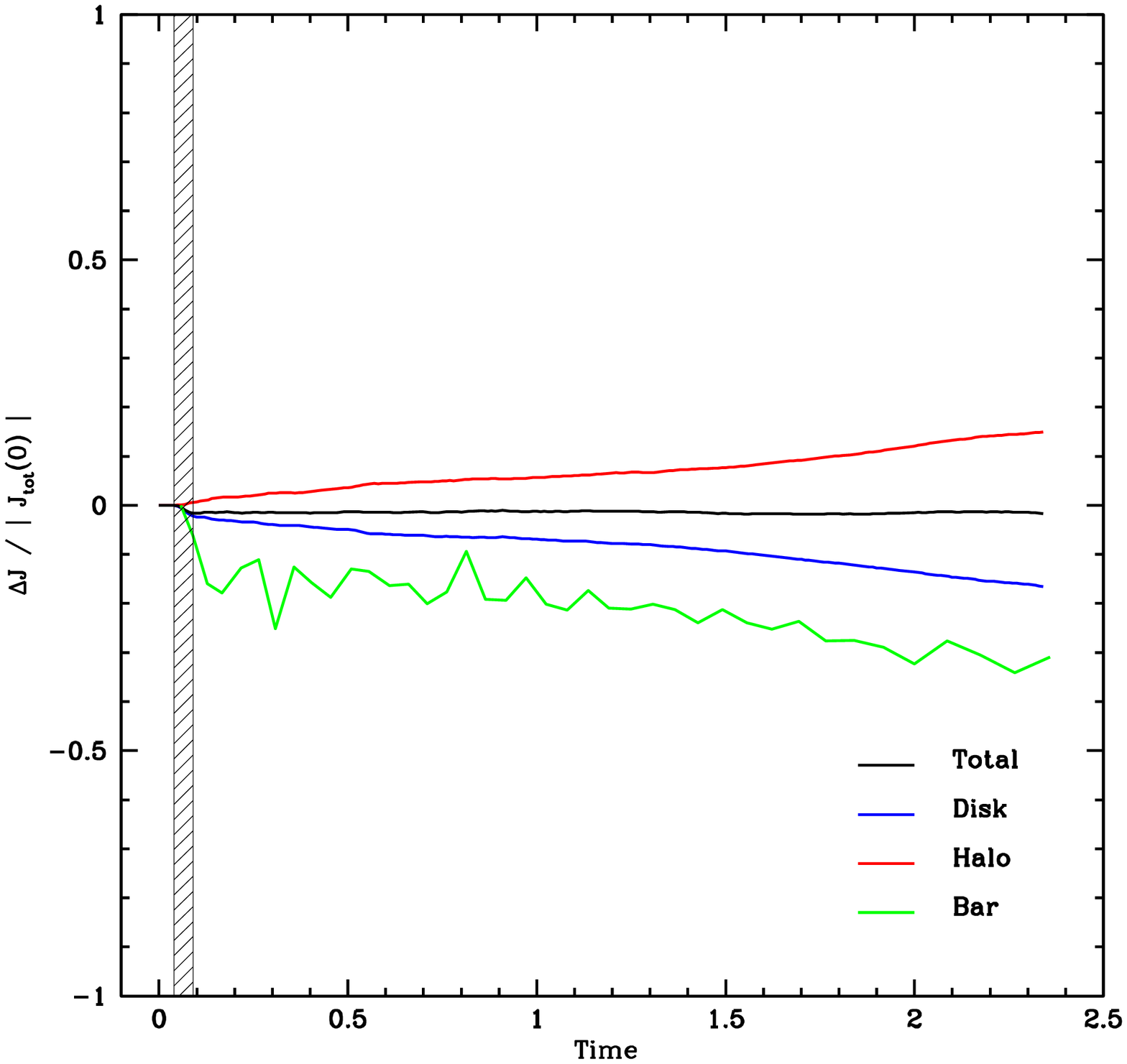}
\caption{The fractional change of angular momentum in each component,
  normalized by the initial total angular momentum for simulation $F_5$.}
\label{fig:angmomrel}
\end{figure}

\begin{figure}[t]
\plotone{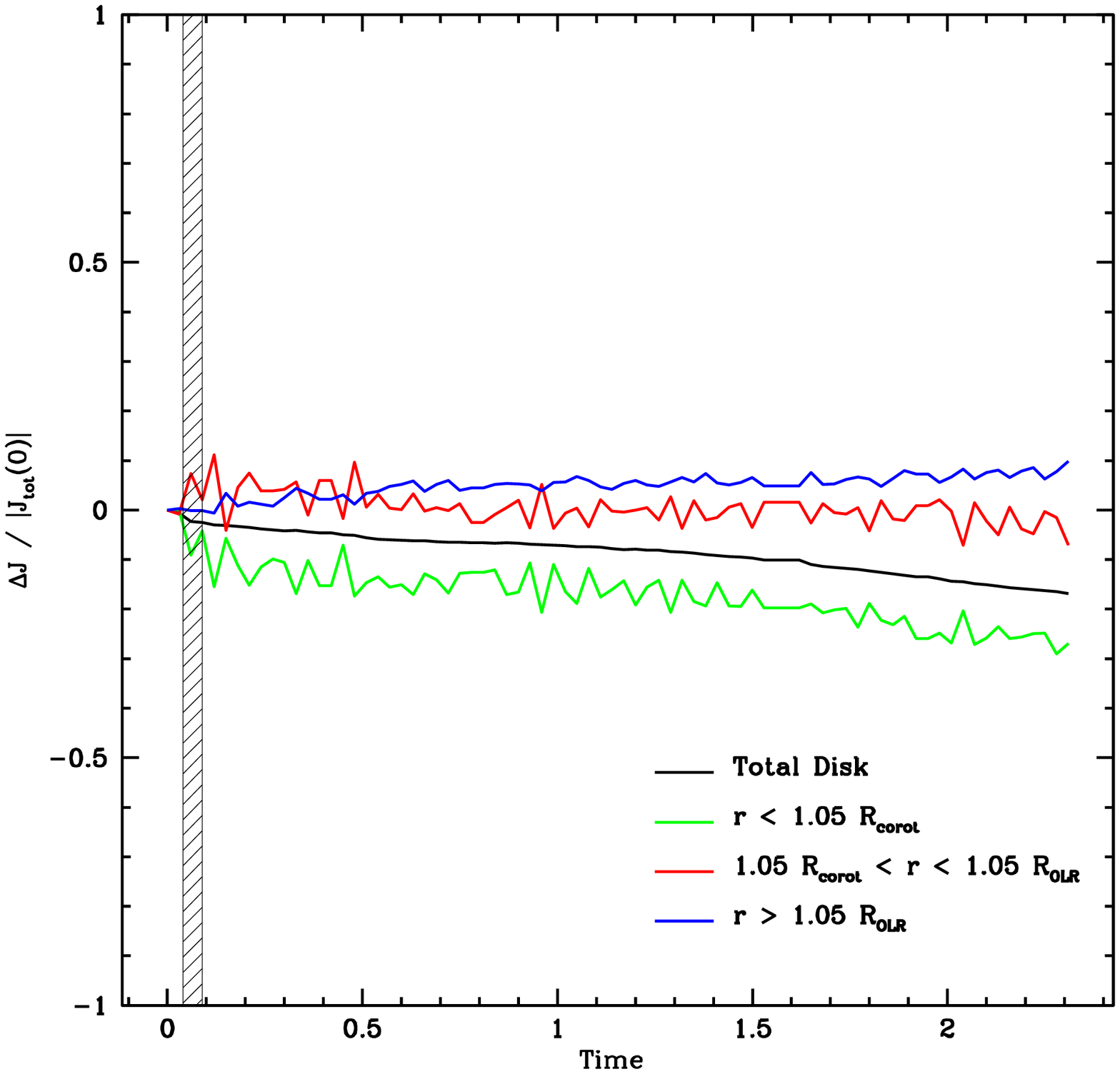}
\caption{The fractional change in angular momentum of the disk for three
  different radial regions, normalized by the total initial angular
  momentum.  The bar length is $R=0.018$ at $T=1.5$, and at this time,
$R_{\rm corot}=0.026$, and $R_{\rm OLR}=0.045$.}
\label{fig:angmomdisk}
\end{figure}

Figure \ref{fig:angmomrel} plots the
angular momentum evolution of the halo, disk, and bar, separately.
Overall, the halo gains $0.0161 J_{\rm init}$ of the initial total angular momentum, $J_{\rm init}$, and the disk loses 
$0.01648 J_{\rm init}$. The total angular momentum of the system is 
conserved to
better than $0.4 \%$ over this time.  
As anticipated from the results of Papers I and II, the bar mediates this
angular momentum transfer, both from the bar to the outer disk and
from the disk to the halo.  The bar loses $0.30 J_{\rm init}$
over the course of the simulation,
with about half being transferred to the remaining disk and half to the
dark halo.  Figure \ref{fig:angmomdisk} shows that most 
of the angular momentum loss within the disk occurs
inside the radius of corotation, $0.28 J_{\rm init}$. However, this is
less than the $0.30 J_{\rm init}$ lost by the bar, which approximately
extends out to the corotation radius. Hence, the material not in the bar
but within corotation gains about $0.02 J_{\rm init}$.  This gain
could be associated with the lengthening and strengthening of the bar.
Also in Figure \ref{fig:angmomdisk}, one sees that between corotation
and the Outer Lindblad Resonance (OLR) the disk loses $0.03 J_{\rm init}$
and beyond OLR gains $0.15 J_{\rm init}$.  Of the 
$0.16 J_{\rm init}$ gain in angular momentum by the dark halo, roughly
$19 \%$ is gained within corotation, $19 \%$ between corotation and OLR,
and $62 \%$ beyond OLR.  This qualitative behavior follows the expectations
of linear theory. Strong resonances in the outer disk and
halo are responsible for most of the angular momentum exchange from the
bar.  In detail, however, things could more complicated. Angular momentum
could be exchanged between different parts of the halo and disk in
addition to those processes implied by the simple angular momentum 
accounting above.  For example, regions
in the halo well beyond OLR actually lose angular momentum, implying
that some additional angular momentum exchange must occur.

\begin{figure}[t]
\plotone{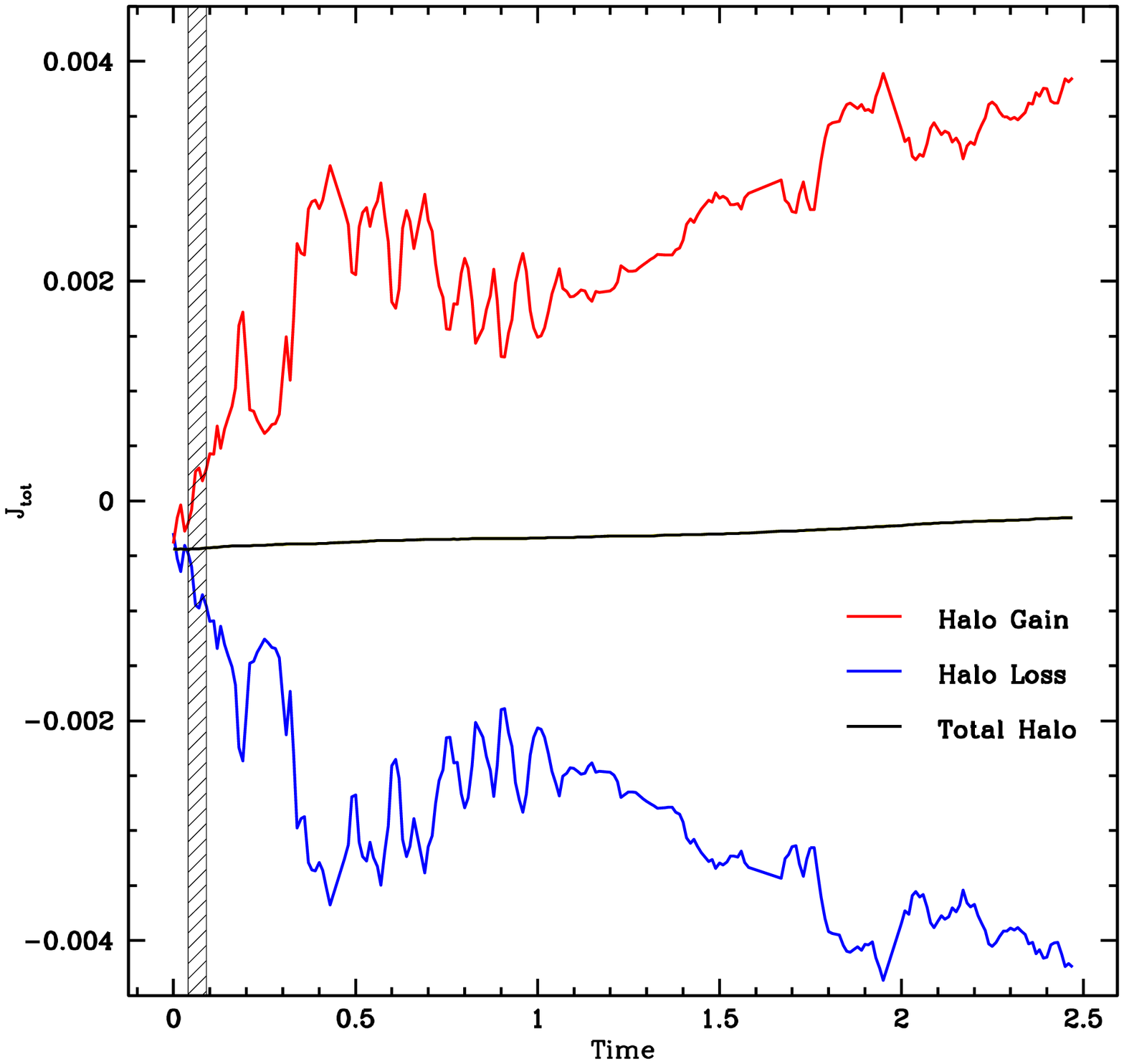}
\caption{The sum of the angular momentum for halo orbits that gain
  (upper) or lose (lower) angular momentum together with the net gain
  (middle) as a function of time.}
\label{fig:angmomhalogain}
\end{figure}

As described in \S2,
near a resonance, the fractional change in the angular momentum of an
individual orbit is first order in the perturbation, and
the sign of the first-order changes are antisymmetric
about the location of the resonance.  The
total angular momentum gains and losses at a given time for individual
orbits are large, but the net change in the
halo's angular momentum is second order, and results from the near
cancellation of this first order effect.
Figure \ref{fig:angmomhalogain} shows both the large
instantaneous gains and losses of angular momentum by individual halo
orbits versus time and the much smaller gains in angular momentum made
by the entire halo, confirming these theoretical ideas.

\subsection{Resonances excited by interactions between the bar and halo}
 
If an N-body system has the correct numerical characteristics to
properly follow the important physical processes, linear perturbation
theory predicts that angular momentum exchange will take place $only$
at confined islands in phase space, the regions that correspond to
halo-bar resonances (as described in \S2 and Paper II). The planar
resonances for a bar rotating with frequency $\Omega_{\rm bar}$ take
the form:
\begin{equation}
l_r \Omega_r + l_{\phi}
\Omega_{\phi} = m \Omega_{\rm bar},
\label{eq:rescond}
\end{equation}
where $\Omega_r$ and $\Omega_{\phi}$ are the radial and azimuthal
orbital frequencies, respectively, $l_r$ and $l_{\phi}$ are integers,
and $m$ is the azimuthal multipole index. Ignoring phase, there are
two non-degenerate actions, or conserved quantities, for a spherical
halo, e.g.  energy $E$ and angular momentum $J$. The resonant
condition therefore describes a curve in $E$--$J$ space.  Contours of
angular momentum change in this space occur in positive and negative
pairs, as orbits either gain or lose angular momentum depending on
their direction of precession just before resonance in the bar's
rotating frame.

\begin{figure}[t]
\plotone{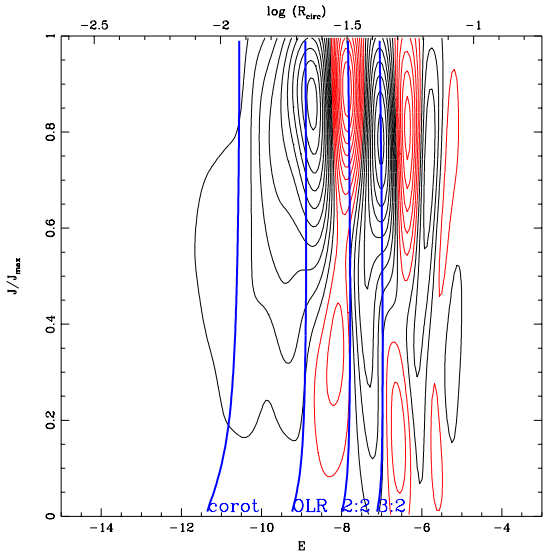}
\caption{The angular momentum, $\Delta J_z$, exchanged between
  $T_1=0.5$ (0.65 Gyr, 1.1 Gyr) and $T_2=0.8$ (1.04 Gyr, 1.75 Gyr) for
  the entire phase-space
  distribution. The horizontal axis plots the total energy
  and the vertical axis plots $\kappa=J/J_{\rm max}(E)$, a measure of
  orbit eccentricity.  The contours represent the difference in
  angular momentum between the two times for each point in phase
  space.  Angular momentum gain (loss) is represented by red (black)
  contours.  The nearly vertical lines are the positions of
  major resonances at time $T_1$, and each are labeled.
  }
\label{fig:resplot}
\end{figure}

\begin{figure}[t]
\plotone{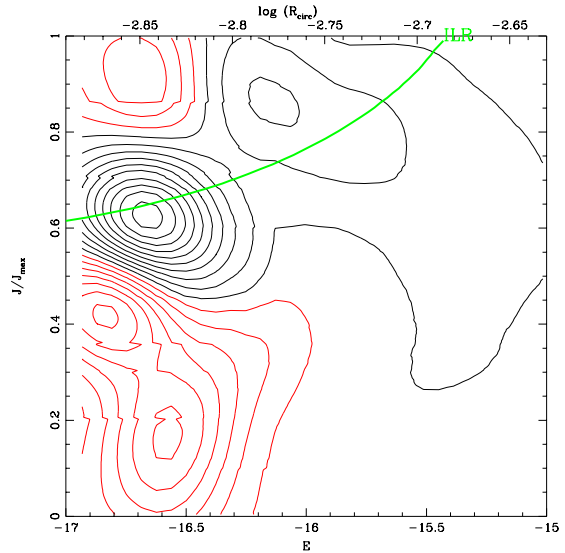}
\caption{The relative angular momentum, $\Delta J_z/J_{\rm tot}$,
  exchanged between $T_1=0.5$ (0.65 Gyr, 1.1 Gyr) and $T_2=0.8$ 
  (1.04 Gyr, 1.75 Gyr) for the
  inner halo, as described in Figure \protect{\ref{fig:resplot}}.
Only the inner region of the halo is plotted; outside this inner region,
the relative angular momentum exchange is negligible.}
\label{fig:resplotrel}
\end{figure}

Figures \ref{fig:resplot}--\ref{fig:resplotrel} show the change in
angular momentum of halo particles between two times, $T_1$ and
$T_2$. The halo phase-space distribution is plotted in the $E$,
$\kappa$ plane, where $E$ is the total energy of the orbit, and
$\kappa = J/J_{\rm circ}$ is a measure of the orbit's eccentricity
($\kappa=0$ and 1 correspond to radial and circular orbits,
respectively). Contours in these figures depict the change in the
z-component of the angular momentum, $\Delta L_z$.  In addition, 
over plotted and labeled are the loci of resonances described by equation
(\ref{eq:rescond}) directly derived from the N-body phase space at
$T_1$. These are typically near vertical lines for our equilibria.

In Figure \ref{fig:resplot}, we plot the evolution between $T_1=0.5$
(0.65 Gyr, 1.1 Gyr) and $T_2=0.8$ (1.04 Gyr, 1.75 Gyr), after the bar
has completed approximately 10 rotations. At this point, the bar has a
stable figure and pattern speed and the resonance signature is
unambiguous, with nearly all the angular momentum exchange in the halo
restricted to easily identifiable low-order resonances with the bar's
rotation.  The main participants in this exchange are the corotation
resonance ($0:2:2$), the Outer Lindblad Resonance ($1:2:2$), and
resonances at $2:2:2$ and $3:2:2$.
 
Figure \ref{fig:resplot} shows the phase-space locations that dominate
the angular momentum lost by the bar and are important for bar
slowing.  However, since halo orbits have differing amounts of angular
momentum, the $\Delta L_z$ contours shown in Figure \ref{fig:resplot} are
not useful for gauging the effects of angular momentum exchange on
individual orbits or on the structure of the halo itself.
In Figure \ref{fig:resplotrel}, we plot the relative
change in angular momentum over the same time period as in Figure
\ref{fig:resplot}.  We quantified the relative angular momentum change
by $\Delta J_z / J_{\rm tot}$, where $J_{\rm tot}$ is the initial
total angular momentum of an orbit.  The largest relative change in
angular momentum takes place within the central region, and the bulk
of this change can be attributed to the Inner Lindblad Resonance
($l_r:l_{\phi}:m = -1:2:2$).  The large relative gain in angular
momentum by these central orbits causes the disruption of the halo
cusp.

\subsection{Density profile evolution}

\begin{figure}[h]
\plotone{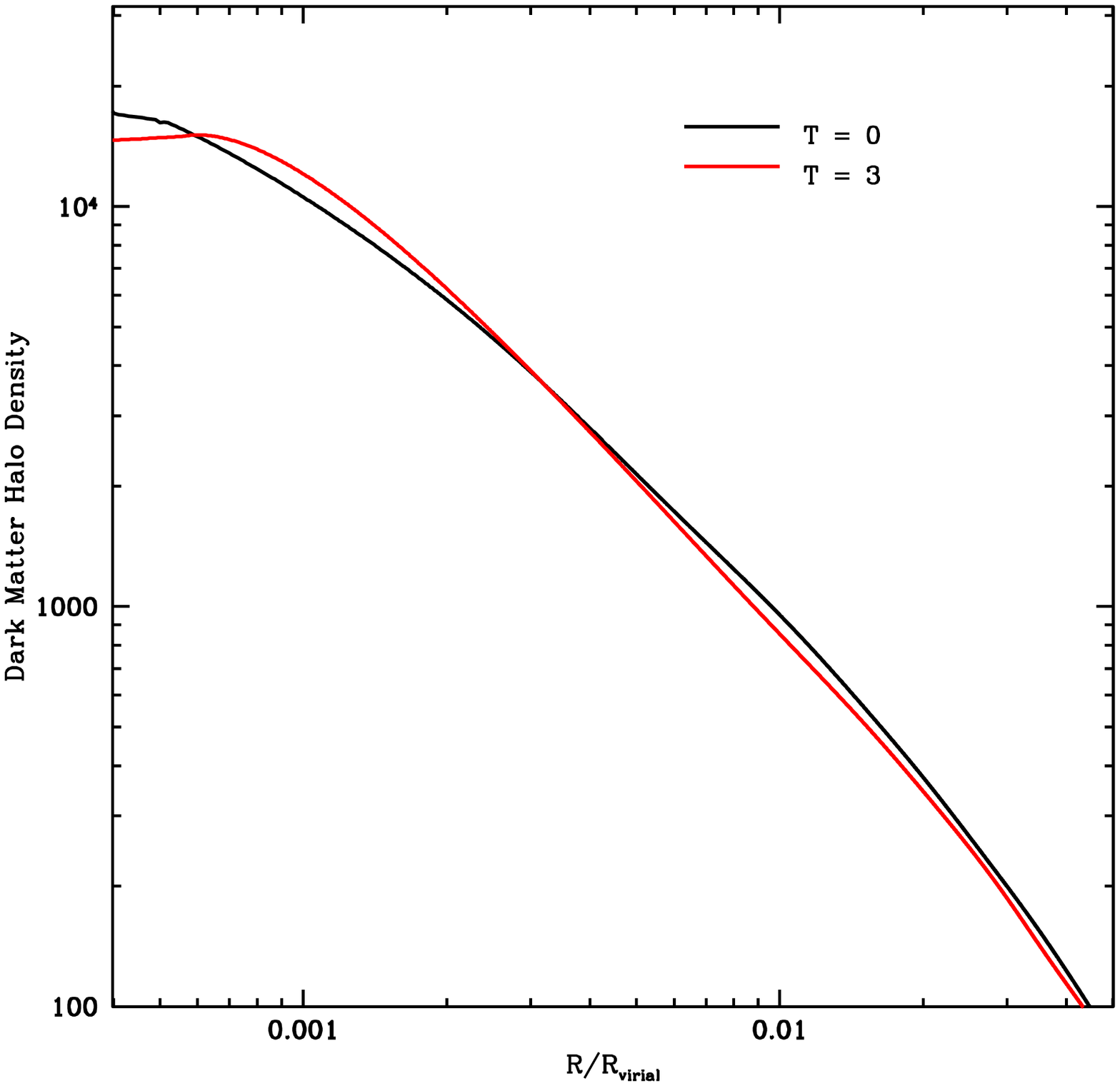}
\caption{Initial and final halo density profiles for a halo with a
  fixed disk potential and 5 million particles ($C_5$).
  }
\label{fig:c5prof}
\end{figure}

\begin{figure}[t]
\plotone{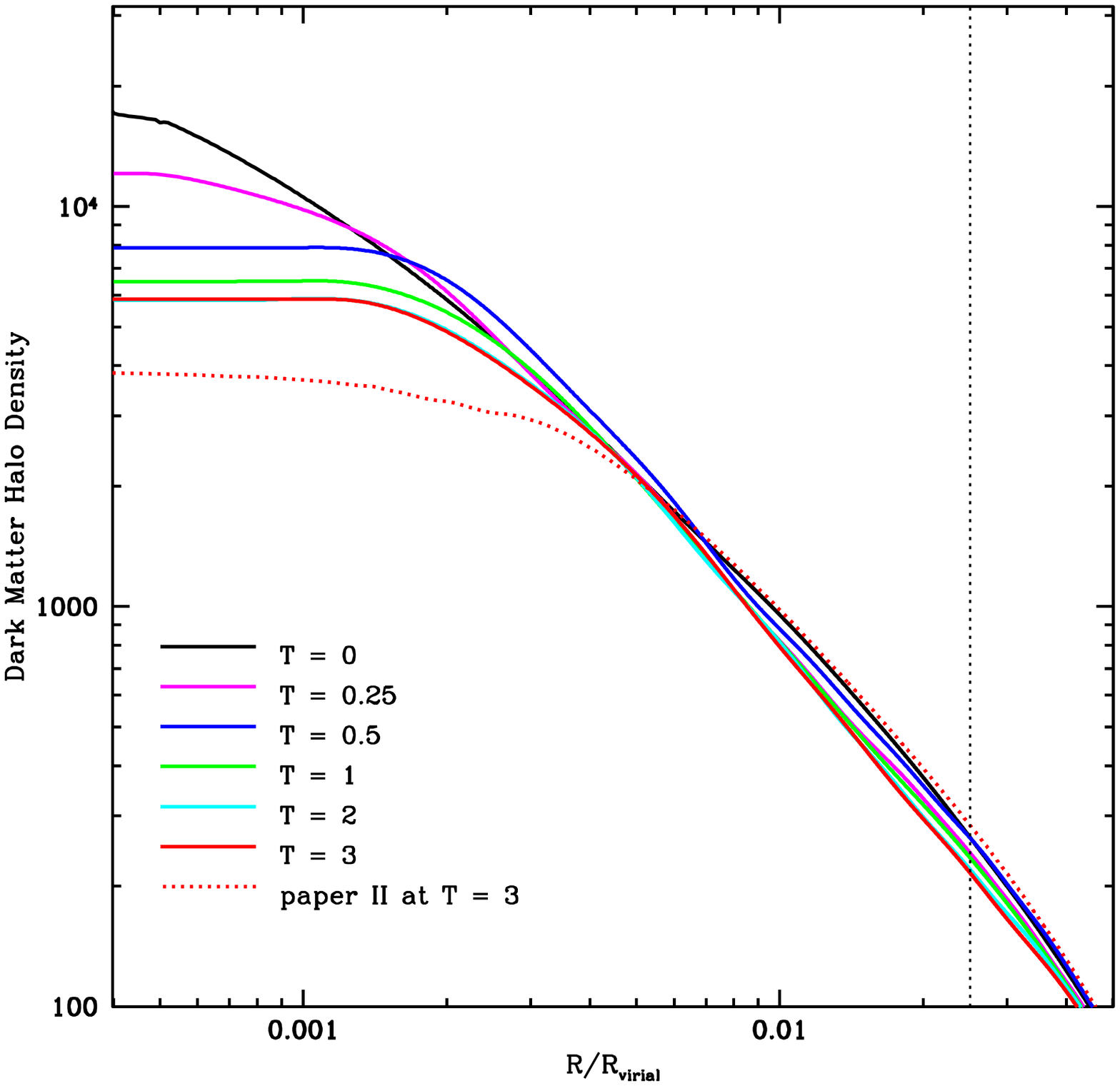}
\caption{Initial and final halo density profiles for the fiducial
experiment $F_5$. See Table \ref{tab:conv} for scaling to physical units. 
The bottom dashed line shows the final state for the idealized 
case from Paper II.
The vertical dashed line shows the final bar radius.
}
\label{fig:f5prof}
\end{figure}

To interpret the effect of the bar on the halo, we must first assess
the stability of the halo profile in the $absence$ of a bar.  Figure
\ref{fig:c5prof} shows the change in the density profile for the 5.5
million particle, fixed disk potential experiment ($C_5$). There is no
appreciable change in the halo density profile over 3 time units 
(3.9 Gyr, 6.60 Gyr).  When a bar is induced via an external quadrupole
($F_5$), the halo density profile changes dramatically, as shown in
Figure \ref{fig:f5prof}.
After $t=1$ (1.3 Gyr, 2.2 Gyr), the halo profile begins to deviate from an
NFW profile at $R=1.67\times10^{-3}$ virial units (97 pc, 500 pc), quickly
flattening to a $\gamma=0$ cusp. At $R=1\times10^{-3}$ (58 pc, 300 pc), the
central halo density has decreased to half its original value. The
bar-induced flattening continues to the end of the simulation at
$t=2.25$ (2.93 Gyr, 4.95 Gyr), producing a core of about 
$R=3\times10^{-3}$ (170 pc, 900 pc).

Our experiments agree with the predictions of linear perturbation
theory and with the idealized simulations of Paper II.  These simulations
investigate the halo evolution driven by the monopole and quadrupole
terms of a rotating ellipsoidal bar with the same size, mass, and 
elongation as the the time average of the F series of experiments, in which
the bar parameters remain fixed.  The final density profile of this idealized simulation
agrees with the fully self-consistent simulation ($F_5$), as plotted in 
Figure \ref{fig:f5prof}, which should be no surprise.  Since the torque
is applied to the halo orbits by the gravitational potential of the
bar, as long as the quadrupole part of this potential in the N-body
simulation is well-represented by the form of the quadrupole used in
Paper II, the evolution and net angular momentum exchange will be
similar. Although the core radii for 
the idealized simulation and $F_5$ {\em are }
similar, the central density of the idealized experiment is about 1.7 times 
smaller than in $F_5$. The increased flattening occurs in the idealized 
run because the quadrupole is fixed over the entire experiment,
while in the fully self-consistent simulation, the quadrupole strength is 
negligible until $T=0.2$ (0.26 Gyr, 0.44 Gyr).
The self-consistent bar structure changes only gradually after
its initial formation, growing slightly more massive and more
elongated as the galaxy evolves (see Fig. \ref{fig:bulkbar}).

\begin{figure}[t]
\plotone{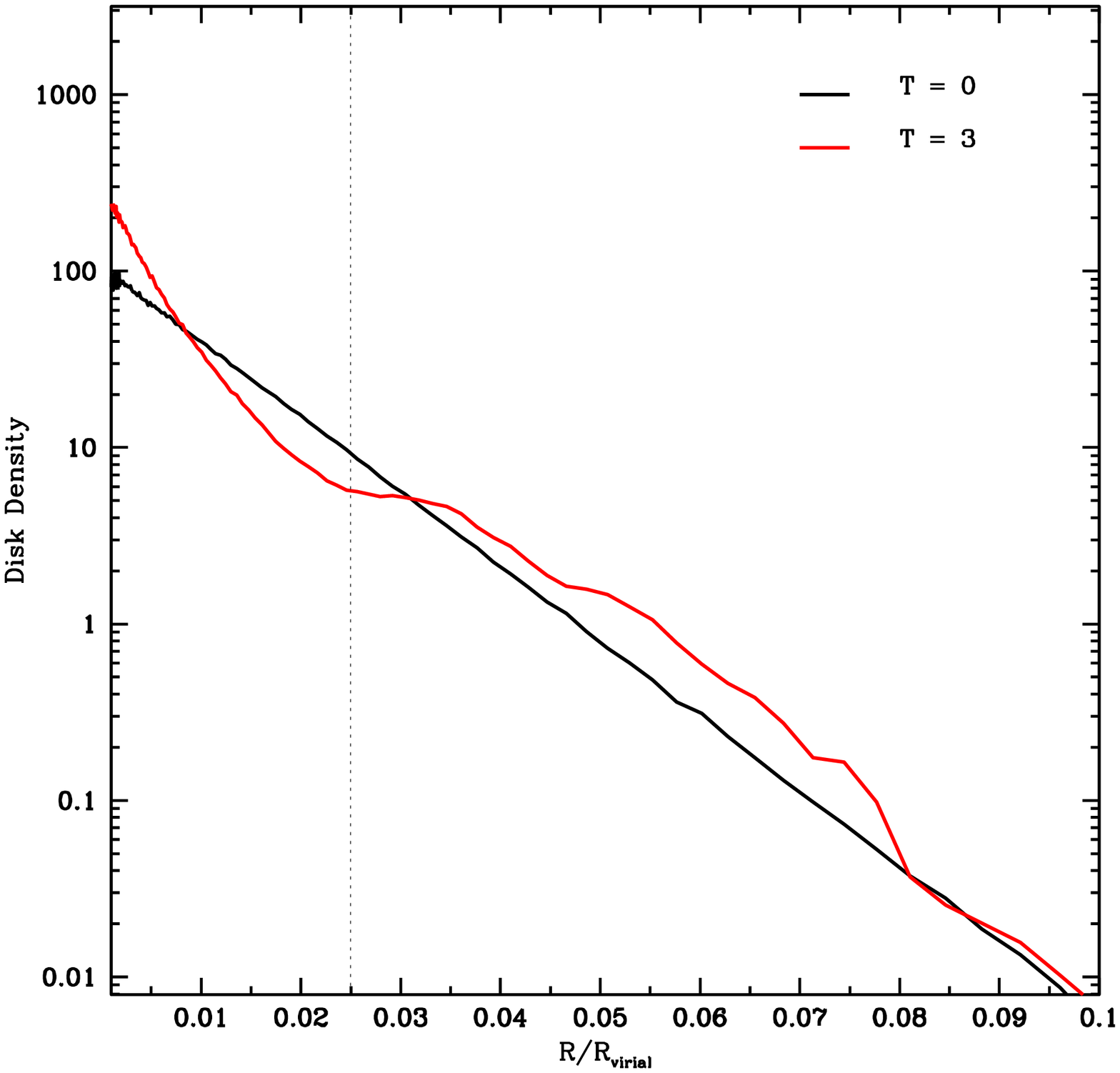}
\caption{Initial and final disk density profiles for the fiducial
experiment $F_5$. The vertical dashed line shows the final bar radius.
}
\label{fig:evoldiskprof}
\end{figure}

The final rotation curve for this simulation, plotted in Figure
\ref{fig:rotcurve}, shows that the disk density profile also evolves,
becoming much more centrally concentrated as it responds to the loss
of angular momentum.  We plot the disk density profile explicitly in
Figure \ref{fig:evoldiskprof}. The inner disk becomes more dense as the
scale length shrinks by $60 \%$ in response to angular momentum loss by the
bar.

\section{Numerical Checks}

Now that we have presented our basic results, it is necessary to determine
their numerical robustness.

\subsection{The effect of the $L=1$ instability}

If the disk were pinned to the initial origin of the simulation, as
in the idealized simulations in Paper I, the
evolution becomes sensitive to $l=1$ instabilities.  As the halo
evolves, random fluctuations produce a small offset between the halo
and the disk.  This offset adds linear momentum to the halo,
further increasing the offset, and leading to a rapidly saturating
instability that artificially amplifies the halo cusp evolution.
Paper I studied bar-induced halo evolution using an NFW halo and an
external quadrupole designed to mimic a rigid rotating bar that was
pinned to the origin throughout the
simulation. This experiment had rapid halo evolution, although some
fraction of it was a consequence of this centering artifact (Sellwood
2002, Paper II). Paper II shows that including a consistent response
of the external quadrupole to the halo removes this artifact and the
rapid halo evolution persists.

\begin{figure}[t]
\plotone{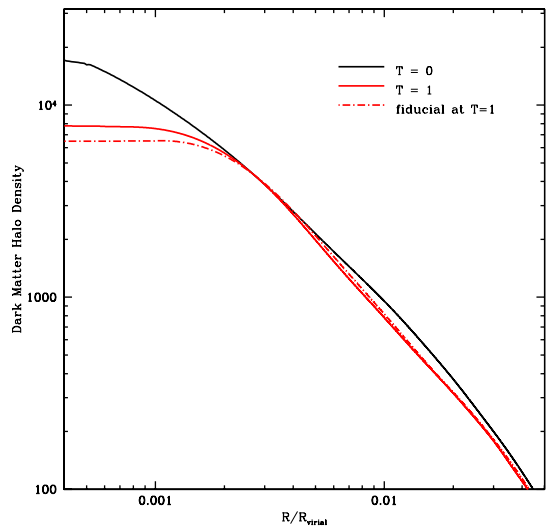}
\caption{Initial and final halo density profiles excluding the $l=1$ part of
the potential expansion. The dashed line shows our fiducial 
($F_5$) final state.}
\label{fig:l1prof}
\end{figure}

The mutual response of the disk and halo in our fully self-consistent
simulations should $automatically$ act to damp this centering
instability.  Each component of the system responds to off-center
density perturbations, which conserves total linear momentum. Hence,
the fully self-consistent adjustment of the halo and disk centers
prevents an $l=1$ instability that artificially arises from a fixed
center. To be certain that the evolution is not affected by
this numerical artifact, we reran our fiducial simulations excluding
the $l=1$ term, i.e. including only $l=0,2,3,4$. If our evolution were
effected by centering, these experiments would show less evolution
than our fiducial runs. Figure \ref{fig:l1prof} demonstrates that the
halo evolution is similar to our fiducial $F_5$ experiment, giving us
confidence that the halo evolution we see is unaffected by this
centering-driven instability.

\subsection{The effect of the external trigger}

\begin{figure}[t]
\plotone{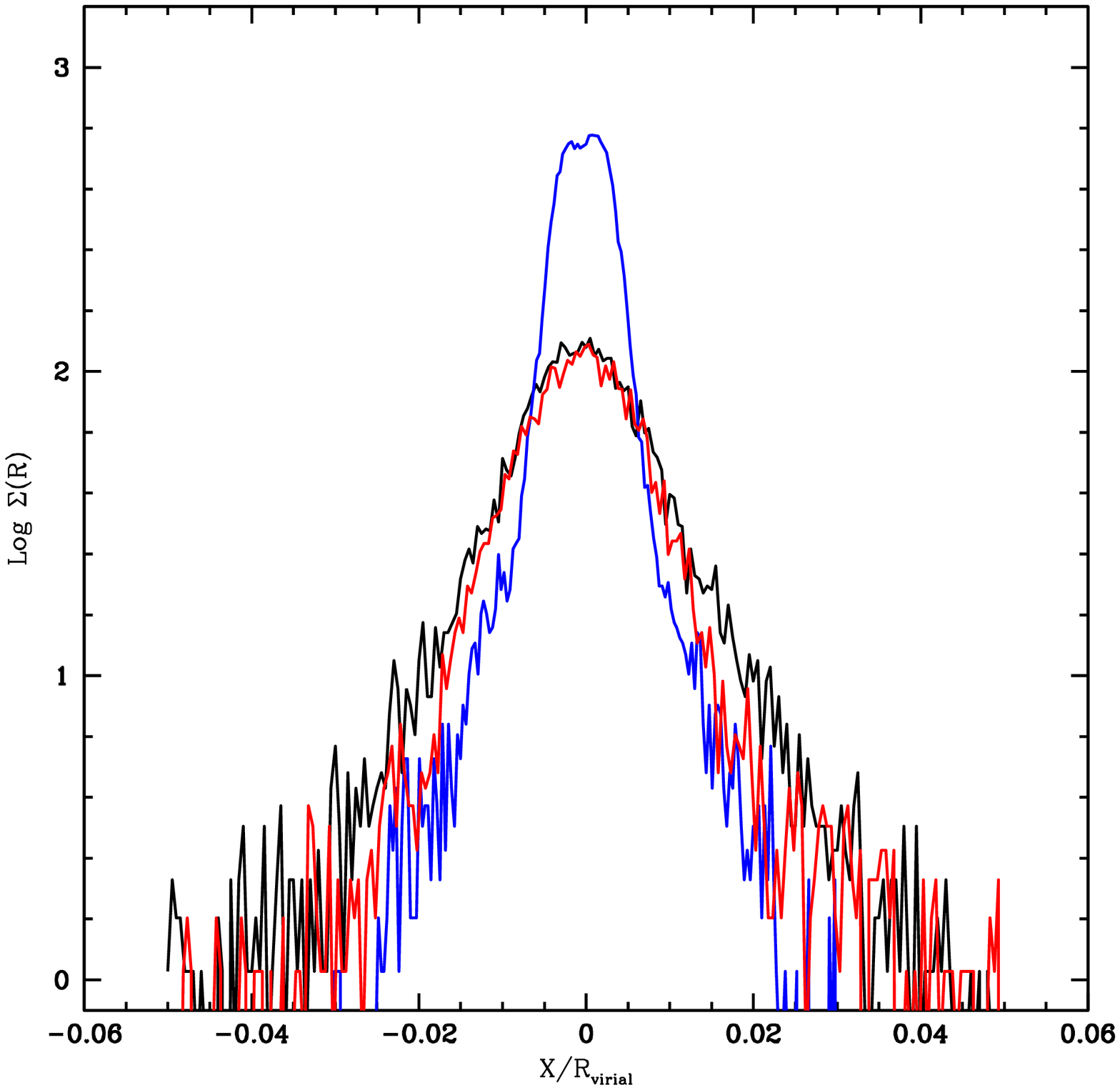}
\caption{Surface density of the disk along the bar major axis for 
  three different bar triggering mechanisms.  The black line corresponds
  to a bar induced by an external quadrupole ($F_5$), and the red line
  corresponds to an instability triggered bar ($I_5$). The blue line 
 represents
the surface density of a bar formed in an adiabatically-grown disk,
as will be discussed in \S5.
  }
\label{fig:dencut}
\end{figure}

\begin{figure}[t]
\plotone{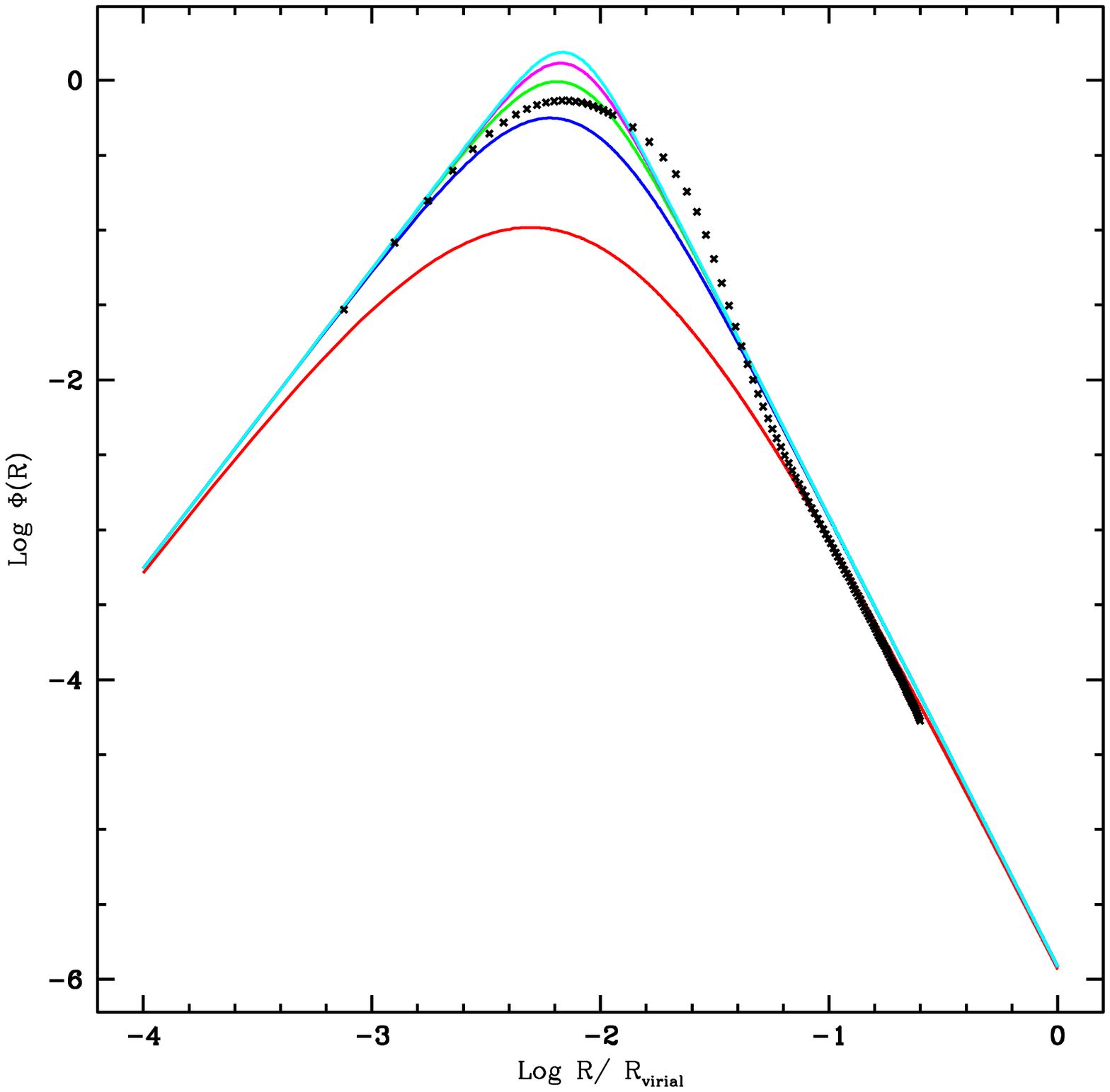}
\caption{The quadrupole component of the gravitational potential of
  the bar. The black points represent the potential derived from the
  $m=2$ component of the disk from our fiducial run ($F_5$) at $T=0.9$
  (1.2 Gyr, 2.0 Gyr).  The
  solid lines correspond to integer values of $\alpha=1,\ldots,5$ for
  the parameterized form of the triggering potential (eq.
  \protect{\ref{eq:quadpot}}).  The bottom curve corresponds to
  $\alpha=1$ and the top curve to $\alpha=5$.
  }
\label{fig:quadpot}
\end{figure}
 
Figure \ref{fig:dencut} compares the projected density profiles of
noise-triggered and external quadrupole-driven bars and shows that the
initial bar length and strength are nearly independent of the
triggering mechanism. Unfortunately, neither bar has surface density
profiles that are as flat as those observed in strongly barred galaxies
(Kormendy 1982, Elmegreen \& Elmegreen 1985); this is a common problem
for bars produced in collisionless N-body simulations 
(Sparke \& Sellwood 1987).  However, a more realistic bar would have
an even greater impact on the dark halo given its larger quadrupole.
Figure \ref{fig:quadpot} compares the
potential in the $m=2$ component of the disk at $t=1$ (1.3 Gyr, 2.2 Gyr), long
after the application of the external trigger, to the analytic
potential provided by the quadrupole in equation (\ref{eq:quadpot})
for integer values of $\alpha$ between 1 and 5. While the form of the
potential used to trigger the bar initially is a centrally shallow
$\alpha=1$, the bar quickly evolves into one with a steeper potential.
Hence, the bar evolution is independent of the precise form
of the triggering potential.  Furthermore, this comparison shows that
the homogeneous ellipsoid is a reasonable choice to model the quadrupole
potential of a self-consistent bar, as we do in Papers I and II.

\begin{figure}[t]
\plotone{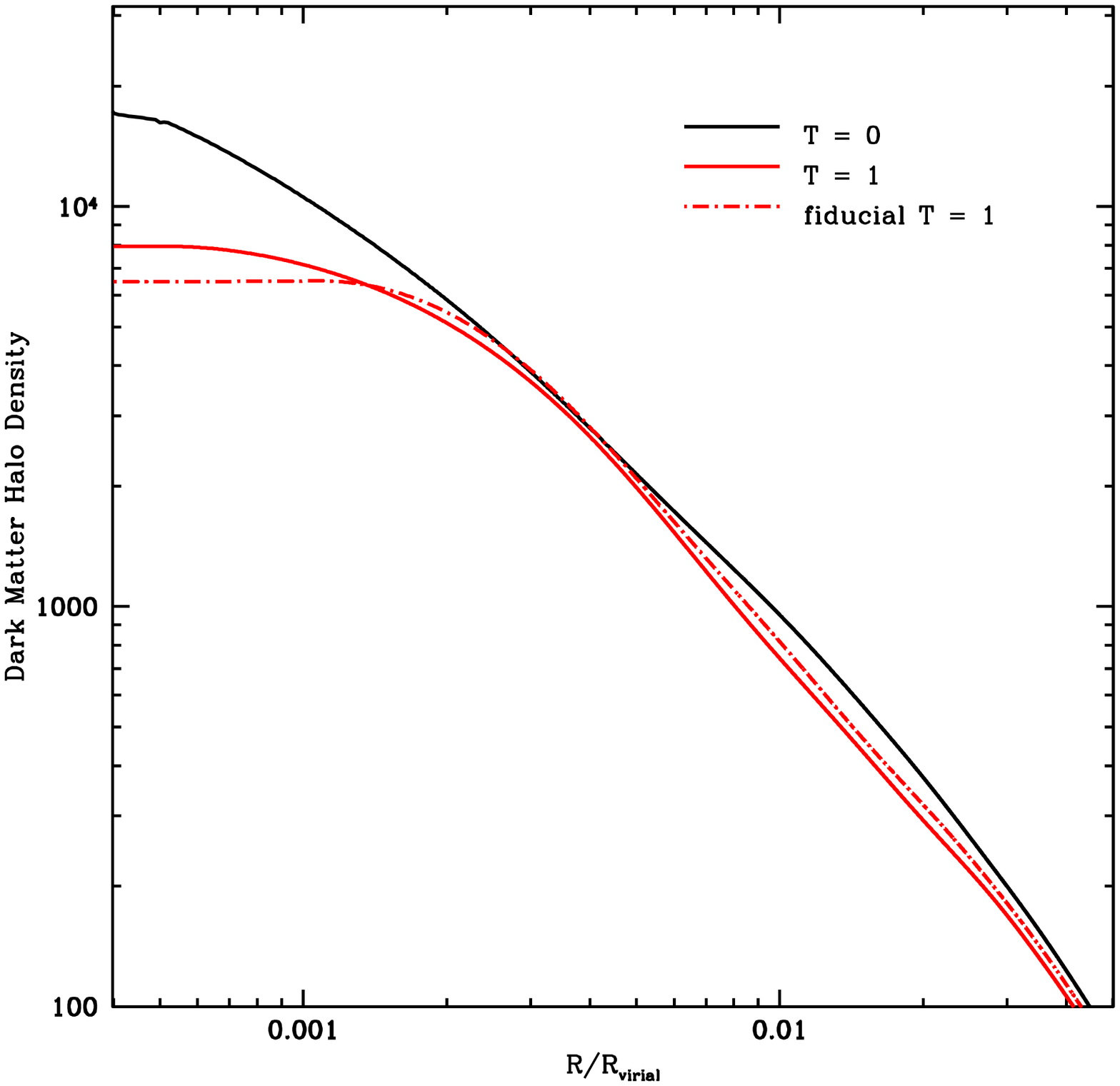}
\caption{Initial and final halo density profiles for the experiment
  that allows the bar to form via internal instabilities
  ($I_5$).  The density evolution is nearly identical to the triggered
  bar ($F_5$), plotted as a dashed curve.
  }
\label{fig:s5prof}
\end{figure}

When we allow the disk to form an approximately scale length-sized bar
through noise-driven instabilities, the halo profile is flattened at
nearly the same rate and to nearly the same radius as the
quadrupole-induced bar (see Fig. \ref{fig:s5prof}).  This similar
behavior from two very different bar-formation mechanisms shows
that the response of the halo does not strongly
depend on the bar triggering mechanism.

\subsection{The effect of particle number}

Even in state-of-the-art N-body simulations, the number of dark-matter
particles is still {\em many} orders of magnitude smaller than those
found in real galaxies.  The errors introduced when the potential is
realized in such a coarse manner can be significant.  This leads to three
possible sources of error: 
there may be an insufficient number of particles to allow the
necessary first-order cancellation to take place near the resonance,
the Poisson noise fluctuations from the
discretely realized phase space can overwhelm the resonant potential,
and the small-scale noise can scatter orbits so that they diffuse past
the resonance, completely obliterating the resonant response.  To
properly resolve the slow evolution of a galaxy with an N-body
experiment, it is critical to use enough particles to ensure that
resonant orbits both exist in the model and are stable for
astrophysically interesting time scales.  Naturally, the minimum
number of particles needed to accurately track the dynamics depends on
both the problem addressed and on the noise characteristics of the
N-body code, so it is difficult to provide a universal particle number
requirement. With current N-body technology, however, long-term
evolution seems to require at least several million particles to fully
resolve low order resonant interactions (Kandrup \& Sideris 2002,
Paper II).  

Here, we empirically determine the critical
particle number required to minimally resolve the resonant
physics important for our problem, and elaborate on the
problems faced when too
few particles are used. In addition, we compare the particle number
requirements derived here to theoretical criteria derived through
perturbation theory analysis and using idealized N-body experiments
in Paper II.

\begin{figure}[t]
\plotone{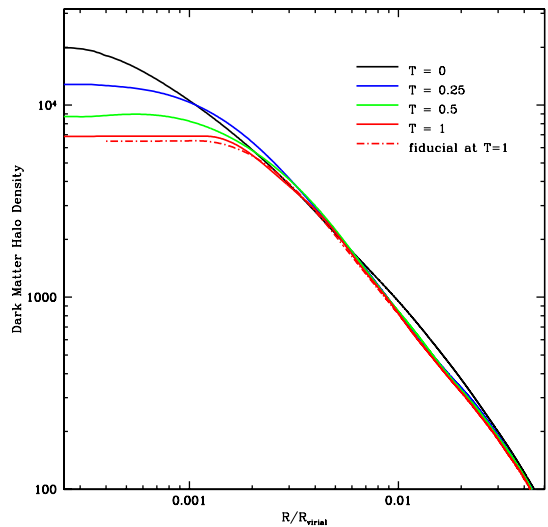}
\caption{Initial and final halo density profiles of our fiducial triggered
bar simulation using 10,000,000 particles ($F_{10}$).  The
  dashed line shows the density profile in the 5 million particle
  simulation ($F_5$) at $T=1.0$ (1.3 Gyr, 2.2 Gyr).
  }
\label{fig:f10prof}
\end{figure}

\begin{figure}[t]
\plotone{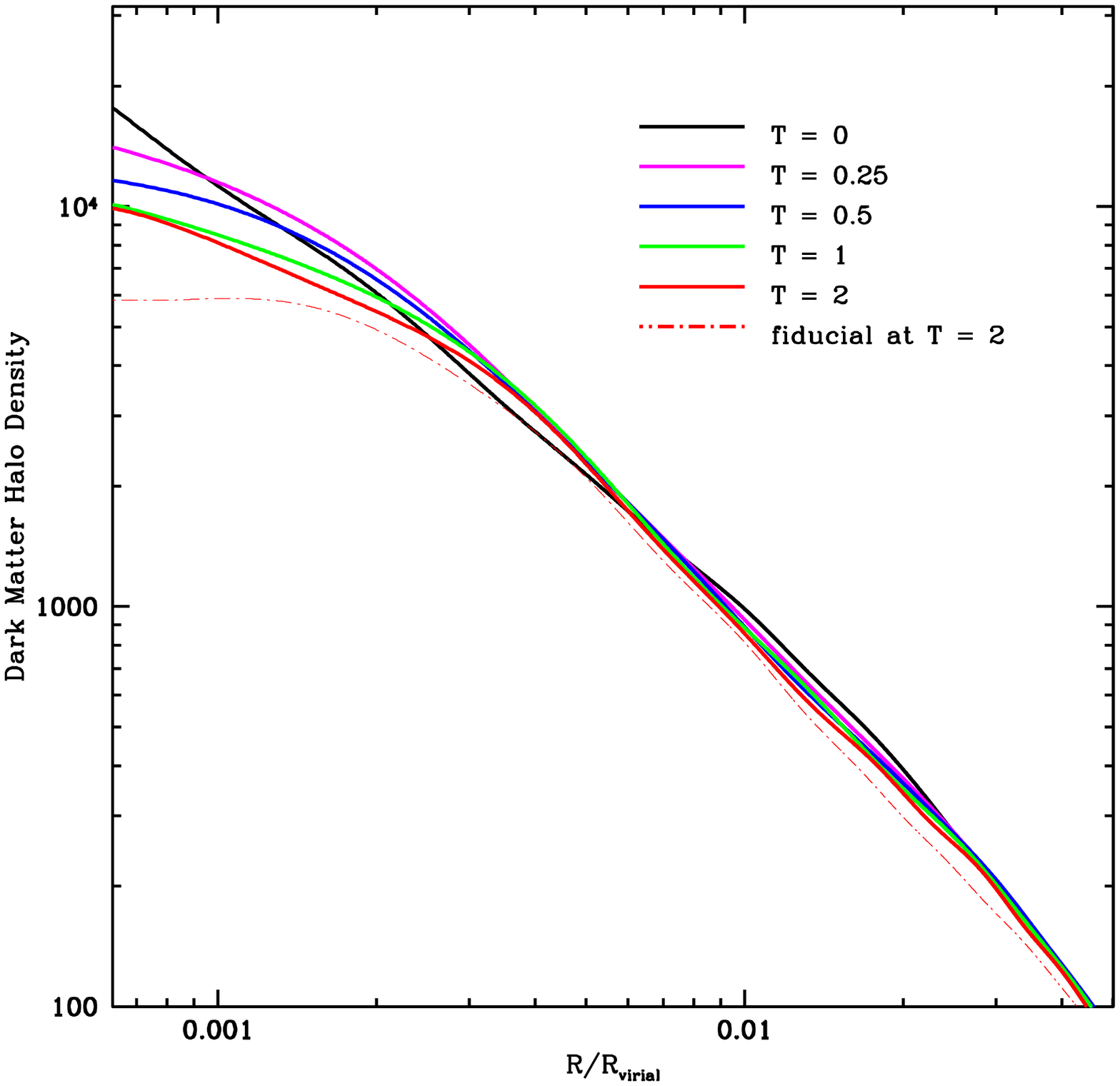}
\caption{Initial and final halo density profiles for the triggered bar
  simulation using 1 million particles ($F_1$).  The dashed line plots
  the $F_5$ density profile at $T=2.05$ (2.7 Gyr, 4.5 Gyr).
  }
\label{fig:f1prof}
\end{figure}

To ensure that the 5.5 million particle experiments are
accurately resolving the dynamics, we compare both the evolution of
1.1 and 10 million particle simulations, $F_1$ and $F_{10}$
respectively, to the fiducial 5.5 million particle run. Experiment
$F_{10}$ yields a nearly identical density evolution to that of our
fiducial experiment, $F_5$, implying that 5.5 million particles are
sufficient for our numerical technique to resolve the dynamics 
of the bar--halo interactions responsible for driving cusp evolution. (Fig. \ref{fig:f10prof}).  Therefore, we consider the halo
evolution observed in our 5.5 million particle simulations to be a
robust estimate of true resonant dynamics between a bar and halo.
Conversely, the relative lack of density evolution in experiment $F_1$,
as evidenced Figure \ref{fig:f1prof}, implies that 1 million equal mass
particles within the virial radius 
is not enough to properly resolve the dynamics.  As we demonstrate
b
elow, the $F_1$ experiment appears to have been plagued both by
insufficient particle number at the crucial resonances and by global
potential fluctuations.  These both act to severely underestimate the
true density evolution of the halo.

As discussed in \S2, any experiment designed to simulate resonant
dynamics must have a well-populated resonance potential. 
This is the fundamental criterion, $f_{\rm crit}$, derived in Paper II.
For example, Runs $F_5$ and
$F_{10}$ show that the ILR is the most important resonance for
increasing the angular momentum of the inner halo. The resonance
potential for the ILR is global and extends from the center of the
halo out to a significant fraction of the bar radius (see Fig. 3 of
Paper II). At least a million particles are required to adequately
sample the outer part of the resonant ILR potential, while
approximately 10 million particles are needed to fully sample the
innermost part of the resonance, which is crucial to the flattening of
the halo cusp (Paper II).  From this consideration alone, it is apparent
that 1
million particle simulations will be unable to follow the halo
evolution physics correctly. In fact, even our $F_{10}$ experiments
only minimally sample the entire ILR. It would be prudent to repeat
all of these simulations with 10 times more particles to better
sample the inner part of the ILR resonance, though this is beyond our
current capabilities.

\begin{figure}[t]
\plotone{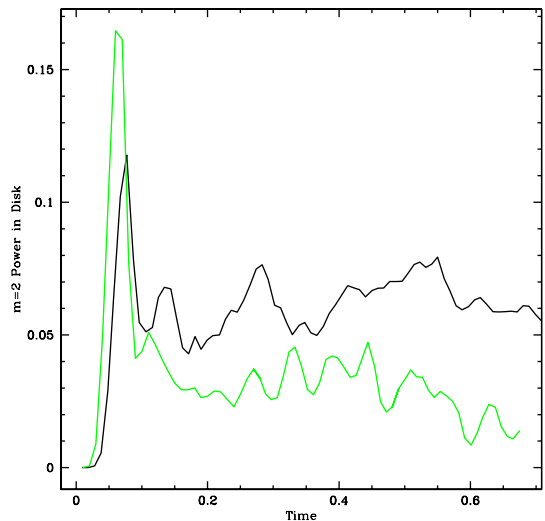}
\caption{The relative power in the $m=2$ component of the disk
  potential for the $F_5$ (black) and $F_1$ (green) simulations.
  After the external trigger is complete at $t=0.1$, the better
  resolved run obtains much more $m=2$ power, while the $m=2$ power in
  the poorly resolved experiment remains relatively constant.}
\label{fig:relpower}
\end{figure}

However, even when the resonance potential is well-sampled, Poisson
noise from the finite-particle realization of the potential can
overwhelm the resonant potential. An appeal to Poisson statistics
demands that the $F_1$ run suffers from larger potential fluctuations.
We can see the effect that these potential fluctuations have on the
quadrupole component of the disk in Figure \ref{fig:relpower}, which
compares the power in the $l=m=2$ harmonic for Runs $F_1$ and $F_5$.
The $F_1$ run features far less power in the $m=2$ mode for both the
disk and the dark matter halo (not plotted).  Despite being triggered
with the same bar strength, the the $F_1$ bar fails to grow as
strongly as in the $F_5$ simulation.  We believe the excess
non-physical noise in the 1 million particle simulation has damped the
quadrupole wake in the halo.

Using the formalism developed in Paper II, we can compare the power in 
the noise to the power in potential from our basis expansion.   
We find that a simulation that resolves up to $n=10$
radial terms requires approximately 3.5 million particles to prevent
Poisson noise from overpowering the signal of the resonant response.
This critical number is similar to the signal-to-noise threshold
derived in Paper II for 10 radial basis terms. Our use of orthogonal
functions to estimate the power at particular scales is no different
than any Fourier-based power spectrum: the maximum number of radial
terms corresponds to smallest spatial scale.  Other potential solvers,
such as direct summation or tree algorithms,
typically resolve many more length scales and consequently have many
more degrees of freedom.  Our estimate based on $n=10$ radial basis terms
will place a lower limit on the particle requirement for these other
methods.

Using these criteria, the reason for the lack of evolution in the $F_1$ 
experiment is clear: it is impossible for 1 million particles to 
simulate the resonant dynamics.

\section{Discussion}

\subsection{Resonances, particle number and comparison with other
  simulations}

The agreement between the simulation results presented here and the
predictions of both linear perturbation theory (Lynden-Bell \& Kalnajs
1972, Tremaine \& Weinberg 1984) and idealized halo--rigid bar N body
simulations (Papers I \& II) provides us with a firm physical
foundation to interpret our results.  We have demonstrated
microscopically that resonances mediate the angular momentum transfer
and drive the structural evolution, a point recently emphasized by 
Athanassoula (2003) in studies pertaining to bar growth.
Differences with other recent studies may be explained either by
differences in the bar shape or by noise effects owing to insufficient
particle number as described in \S2 and \S5.3, or both.

In these recent studies, the halo is represented as an approximation
of a smooth system, yet approximately $10 \%$ of the mass within a dark
matter halo is contained within clumps between $10^8$ and $10^{10}
M_\odot$ (Moore et al. 1999b, Font et al. 2001).  This substructure
noise spectrum provides large, low frequency fluctuations and Weinberg
(2001ab) shows that resonance driven evolution does occur in the
presence of this real astronomical noise.  In contrast, {\em N-body}
discreteness noise generates high spatial frequency fluctuations that
drive evolution by local diffusion.  A sufficiently rapid drift in
orbital frequencies decouples orbits from any resonances and
diminishes or eliminates the torque.  Therefore, noise on
interparticle scales has very different consequences than large-scale
dark matter substructure.  Although astrophysical noise is most
certainly an important contribution to galaxy evolution, astrophysical
and numerical noise sources have different effects on the evolution
studied here. Astrophysical noise causes the location of resonances to
slowly change while numerical noise causes them to disappear.

We have presented two particle number criteria necessary for
resonances to be effective (see Paper II for details) and have tested
them empirically. For equal mass halo particles, at least 5 million
particles are required within the virial radius to fully resolve the
resonant physics. A third criterion requires a sufficient particle
number so that an orbit does not diffuse away from resonances; we
defer development of this to a later paper.  This orbit diffusion
criterion is not critical for our SCF N-body method, but might prove
to be the most restrictive for direct summation, tree or grid based
techniques.

Spatially adaptive codes such as direct summation, multigrid or tree
codes can resolve acceleration from structures on many scales.  This is
a great advantage when simulating large-scale structure, but unfortunately
also allows noise to appear at all scales.  At small scales, 
this noise is
dominated by finite particle sampling rather than astronomical
structure.  In contrast, our SCF expansion method selectively
filters the small Poisson-dominated scales, reducing the small scale noise
by at least an order of magnitude. Even though
the small scale noise is reduced, our choice of expansion parameters
still allows us to fully resolve the resonances themselves. Any further
increase in spatial resolution runs the risk of adding to the numerical
noise and removing the resonances.

Even though the resonances responsible for slowing the bar occur at 
large radii and are, therefore,
easier to accurately resolve, we suspect
that the noise characteristics of many widely-used potential solvers
have suppressed the torque in those studies that find little
evidence for bar slowing.  The numerical criteria for accurately evolving
the cusp are more stringent, both because the resonances responsible
for the cusp evolution extend to much smaller radii where there are fewer
particles and because the resonance responsible has a much weaker, 
more shallow potential. We suspect
that many studies that fail to find cusp evolution are not adequately
resolving the central resonances.

We show in Paper II that the
bar slow down rate and the change in the halo density profile
depend sensitively on the
shape, amplitude, and size of the bar.  Bars with a smaller amplitude
take longer to slow, have less effect on the cusp, and are more
numerically difficult to evolve correctly due to the shallower
resonance potential.  Although the size of the bar does not greatly effect
the slow down rate, a
highly centrally concentrated bar has a larger fraction of the power
in the m=0 mode, and with less m=2 power it is less capable of 
driving cusp evolution.  Smaller bars are also more numerically difficult
to evolve accurately since the resonances lie at smaller radii.
These numerical and physical effects combine to make simulating the
evolution of galaxies with small, weak bars particularly subtle.

For example,
VK perform simulations of bars using the
Adaptive Refinement Tree (ART) potential solver (Kratzov, Klypin \&
Khokhlov 1997), allowing the bar to form through local instabilities
and following its evolution.  They find that a bar forms with a radius
about equal to that where the rotation curve stops rising, that the
bar does not slow appreciably, and that the central dark matter cusp is
not removed.  
All of their differences with our results can be easily explained by their
inability to properly follow the resonant dynamics.  It prevents
their bar from secularly growing and it remains small.
Since resonances mediate the
angular momentum exchange between the halo and the bar, the bar will not
slow and the cusp will not evolve if the resonances are removed.
Their simulation with the largest particle number, $N=3.55\times 10^{6}$
particles within the virial radius,
satisfies our necessary criteria for populating the resonances in
phase space.   This number of particles is also enough to suppress
fluctuations on the scale of the resonant potential, though more particles
are required due to their
higher spatial resolution. Their multimass technique
partly but not entirely mitigates the problem by concentrating more
particles in the central region, making their particle number equivalent to 
$9.5\times 10^{6}$.

However, this and any mesh-based N-body technique suffers from
numerical artifacts that artificially decrease the 2-body relaxation
time scale (Gneidin \& Bertschinger 1996, Dehnen 2001). The short
diffusion time scale may scatter orbits too rapidly for resonant
angular momentum coupling to take place. Although we cannot determine
the diffusion time scale in the VK experiments without their initial
conditions and code, previous comparisons of the noise characteristics
of tree and expansion codes indicate that about 2-4 times more
particles are required for a tree code to achieve the same diffusion
time scale as an equally-smoothed SCF code (Hernquist \& Ostriker
1992, Barnes 1997), implying that a 20 million particle simulation would
be required to achieve stable orbits in the halo for a standard grid based
code with our spatial resolution.  VK also have extremely high spatial
resolution, which
further exacerbates the small scale noise problem and would require even
more particles, as discussed in the previous paragraph.

Furthermore, the orbit diffusion problem is even more acute in the
ART code.
Adaptive refinement serves to better resolve regions of
high density, making this a excellent technique to model the halo 
cusp.  However, if
the mesh used to determine the potential is not adjusted often enough
to respond to the changing system, as in the simulations of VK, the
particles can experience unnaturally large accelerations across mesh
boundaries that are unrelated to the true potential (Jessop et al
1994, Anninos, Norman, \& Clarke 1994, Kravtsov, Klypin, \& Khokhlov
1997).  VK choose to regrid only after many particles enter a grid cell
to reduce run time.  Particles
within a grid cell are free to drift relative to one another until
regridding occurs.  If particles have drifted close together, when regridding
occurs, they will suffer artificially large Fermi accelerations that
will scatter the orbits and pump energy into the system.  This will greatly
increase the orbit diffusion and swamp the resonant bar-halo interactions.
Hence, the
lack of inner halo evolution and the comparatively smaller overall
bar--halo torque reported by VK might be partly caused by their
technique.

Sellwood (2002) also studied the formation and evolution of bars that formed
through local instabilities using a grid based approach.
He forms a bar that is both weaker and more centrally concentrated than those
in our simulations (e.g. $I_5$). It slows, but does not affect the
central dark
matter cusp.  His largest simulations contain 21 million particles,
enough to satisfy both of our criteria for properly evolving
resonant dynamics but perhaps not enough to sufficiently reduce orbit
diffusion.  He also uses different initial conditions than either us
or VK.  He uses a Hernquist profile instead of an NFW profile and,
more importantly, instead of starting with an equilibrium model he
grows the disk adiabatically.

To investigate the
effect of Sellwood's disk growth procedure, we adiabatically grew an
$I_5$ disk in an initially $I_5$ halo (see Table \ref{tab:barparams}),
allowing the bar to form by instability.  The surface density profile
of the resulting bar is sharply centrally peaked, shown in 
Figure \ref{fig:dencut},
as it also appears to be in Figure 6 of Sellwood (2002).
The self-consistent bar in the Sellwood study is unlike both
the fiducial bar in our work and observations of most strong bars,
which have constant surface density profiles (Kormendy 1982, Elmegreen
et al. 1996). 
Such a highly centrally concentrated bar is less capable of
driving cusp evolution.
In addition, the bar appears to be much weaker than in $I_5$.
Our adiabatically formed disk is 1.5 times hotter than the one in our
equilibrium initial conditions, and Sellwood's use of a
rather low disk to dark halo mass ratio of 0.05 further
raises the Q value; local instability triggered
bars in hot disks are smaller and weaker. Since
halo evolution scales with the bar mass, we believe the lack of halo 
evolution is likely a consequence of an anomalously weak quadrupole.
Finally, since it is more numerically challenging to evolve
smaller, weaker bars,
Sellwood's simulation might not
be able to resolve the inner resonances responsible for the cusp evolution
but still be able to resolve those responsible for slowing the bar.

Athanassoula (2003) presents self-consistent N-body simulations
of a forming and evolving bar using a direct summation approach implemented
with GRAPE.  Her bars form through local instabilities and
many of her results are consistent with our findings.
Most of her bars slow and she shows that 
angular momentum transfer to the halo, mediated by resonances,
determines the slow down rate.
Although not discussed in the text, her Figures 8 and 12 clearly indicate a
reduction in the central dark matter density.
Her halo profile is an 
isothermal sphere with a core truncated at about 1/9 the virial
radius.  In her typical experiment,
she uses 2 million particles, which would be equivalent to about 
10 million particles if she simulated an NFW profile that extended
to the virial radius.  In addition, she uses a multimass technique that
further increases her effective particle number to at least 25 million.
Once again, this is enough particles to satisfy both of our
particle number criteria for properly evolving the
resonant dynamics but perhaps not enough to sufficiently reduce orbit
diffusion in a direct summation code.

However, in her run that best
matches our fiducial simulation, MHH2, the change in the bar pattern speed
was about 3.5 times smaller than ours.
This run features a more extended halo,
based on set of initial conditions that add a
second component with larger extent and a larger core. She finds that
bar slow down rate decreases with increasing halo extent and attributes this
trend to a higher velocity dispersion in the more extended halos.  Although
this could in theory explain her results, Athanassoula's experiment
does not represent a fixed profile successively truncated at smaller
radii.  She increases the extent of her initial halo by adding a
component with a larger core radius.  This new profile increases $M(<r)$ not
only in the outer parts but
at all radii, which not only changes the velocity dispersion, but also
changes the dark-matter gravitational potential and, hence, the orbital
frequency distribution near the bar.  In Paper II
we perform simulations using our idealized bar model
with NFW profiles truncated from $0.3 R_{\rm vir}$
to $2.0 R_{\rm vir}$.  In all cases, the bars slow down at roughly the same
rate and the velocity dispersions at the bar radius do not measurably change.
The velocity dispersion only changes significantly
near the truncation radius.

Moreover, Athanassoula's extended halos modify the stability of
the disk, i.e.  the dispersion relation, by contributing additional
gravitational support, thereby modifying the length and shape of the
forming bar.  In fact, the bars that formed in her
multicomponent halo systems are smaller and weaker than those in the
single component halos (Athanassoula 2003, her Fig. 8).  The shape of
the bar's quadrupole critically determines the strength of the
coupling with the halo as a function of radius (see Paper II).
This both explains the differing torque estimates and
suggests that the trends reported in Athanassoula (2003) may depend
sensitively on both the bar shape and halo profile.  In addition,
smaller bars require a larger number of particles to adequately
resolve the resonances and too few
particles could systematically diminish the torque.

Bars form through the secular growth of features either triggered by
local instabilities or externally, e.g. by satellite encounters.
Resonant angular momentum exchange with both
the dark halo and the outer disk facilitates this secular growth
(Athanassoula 2003). 
Hence, to accurately simulate bar growth one must properly simulate resonant
dynamical processes, requiring the simulation to meet all the numerical
criteria discussed above.
The size and strength
of bars that form through local instabilities depend on the
stability of the disk, i.e. disks with lower Toomre Q values will form
stronger bars than those with larger Q values.  Since the formation of
a local instability triggered bar depends on the existence of an ILR,
the initial bar sizes are restricted to the
rising part of the rotation curve.  However, secular growth
can further
increase the bar size well into the flat part of the rotation curve.
In the simulations of Athanassoula (2003), bars form through local 
instabilities that
are two to four times larger than the radius where the rotation curve
becomes flat. In contrast, when bars form due to an external trigger, even
their initial size is not restricted to the rising part of the rotation
curve.  The large imposed
quadrupole potential from the tidal trigger overwhelms the ``donkey
star'' growth criterion (Lynden-Bell \& Kalnajs 1972) based on
the underlying potential alone.  Similarly, a tidal trigger nullifies
formation arguments based on swing amplified instabilities (e.g. Mayer
\& Wadsley 2003).
To demonstrate this, we perform a simulation ($B_5$) where we
externally trigger and form a large stable bar using our fiducial 
simulation ($F_5$) parameters, but with a quadrupole scale length that
is twice as large.
The bar forms with a radius of $R=0.04$ (2.3 kpc, 12 kpc) after $t=1.5$
(2.0 Gyr, 3.3 Gyr), as shown in Figure \ref{fig:bigbar}.
Remember that the disk had an initial scale length of 
$R=0.01$ (580 pc, 3 kpc) and that the rotation became flat at 
$R\sim 0.01$ (580 pc, 3 kpc).

\subsection{Comparison with observations and $\Lambda$CDM galaxy
  formation scenarios}

Our N-body experiments demonstrate that an initially corotating,
scale-length sized bar will generate a core in the density profile
that extends out as far as $R=0.003$ (170 pc, 900 pc).  Current rotation
curve decompositions of high surface brightness spirals (Salucci \&
Burkert 2000) suggest halo core radii of at least $R=0.04$ (2.3 kpc, 12
kpc). The dynamics of the gas and stellar components may overwhelm 
the dark matter signal in these systems, however.  Low
surface brightness galaxies (LSBs) are thought to be dark matter
dominated and can, therefore, be used to obtain a cleaner dark matter
signal from the rotation curve.
Typical halo core radii for low-surface-brightness galaxies
range from 0.001 (60 pc, 300 pc) to 0.02 (1.2 kpc, 6 kpc).
Hence scale-length-sized bars, like those studied here, cannot
produce cores this large in cuspy $\Lambda$CDM density profiles. 

\begin{figure}[t]
\epsfig{file=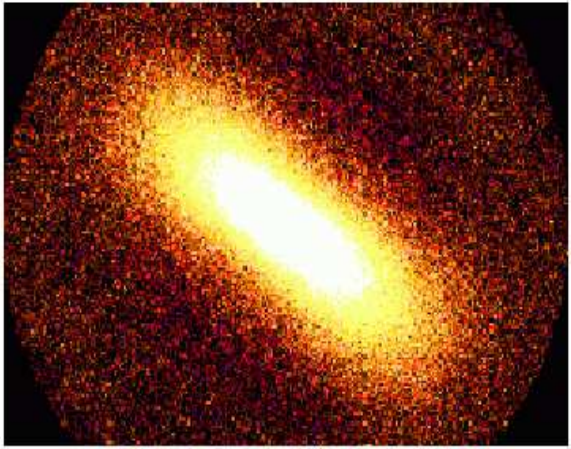,height=200 pt,width=250 pt}
\caption{Face on view of a $R=0.04$ (2.3 kpc, 12 kpc) externally triggered
bar ($B_5$) after $t=1.5$ (2 Gyr, 3.3 Gyr) in a disk with an initial
scale length of $R=0.01$ (580 pc, 3 kpc).}
\label{fig:bigbar}
\end{figure}

However, the lengths of {\em tidally-triggered} bars can be over
10 times larger than those in our
fiducial simulations.  As we have discussed in the previous section, we form a bar four times our fiducial length through an external trigger
(Fig. \ref{fig:bigbar}). Since linear perturbation theory suggests that
the destruction of the halo cusp scales with the size of the
perturbing quadrupole (Paper I, Paper II), we expect that larger primordial
bars will generate proportionately larger cores.  

A second possible solution invokes multiple
epochs of bar formation and destruction during the hierarchical assembly
of the galaxy, which is characterized by relatively quiescent periods punctuated
by mergers.  Since conservation of phase space density suggests that the 
merger of two halos with a core also
forms a halo with a core, the cores will persist as long as one or
more bar phases occur sometime during the hierarchical assembly process.
It is plausible that a new disk equilibrium is established after each 
merger event and that the quadrupole of the merger remnant will externally
trigger a large bar.  Minor mergers will also trigger bars without
disrupting the disk.  A larger core at the end
of each merger-punctuated epoch facilitates the formation of a larger
bar and subsequently a larger core.  Taken more generally, our
bar--halo mechanism predicts an intrinsic dispersion in galaxy properties
owing to differences in evolutionary history: the present-day
morphology will depend on whether or not a large bar was triggered,
whether or not a standard bar is excited by astrophysical
noise sources, and the overall merger history.  This 
dependence can lead to galaxies with varying degrees of cusp
flattening and disk scale-length evolution.

Since most of the evidence against central dark matter cusps in galaxies
concerns low surface brightness dwarfs,
one might not expect strong bars to form in such systems and, hence,
for the mechanism proposed here not to have much relevance.  However,
the same analysis used to indicate the lack of a central dark matter  
cusp also shows that these low surface brightness galaxies are three or
four times more baryon deficient than normal
galaxies (Van den Bosch \& Swaters 2001).  If a strong bar forms in a
gas rich disk of a dwarf galaxy, much of the gas will lose substantial
amounts of angular momentum, be driven towards the center,
undergo a strong starburst and, due to the shallow potential well
of the dwarf galaxy, much of the gas could be expelled as a
supernova-driven wind (Dekel \& Silk 1986).  The work done on the galaxy
during this process will cause further expansion of the core.
The remaining galaxy
would be one of low surface brightness, possessing a core in its dark 
matter distribution.  Our bar mechanism, therefore, not only provides
a natural explanation for the existence of dark matter cores, but for
the existence of low surface brightness dwarfs as well; those dwarf
galaxies that had the strongest bars will have the largest cores and the
lowest surface brightness.  Furthermore, Holley-Bockelmann, Katz, \&
Weinberg (2003) show that even low surface galaxies can form bars and
remove their dark matter cusps.

Regardless of the scenario, we have set a lower limit to the
size of the core generated by early bar--halo interactions.  This
lower limit already has implications that are astrophysically
relevant.  For example, Gondolo \& Silk (1999) have argued that the cusps
of dark matter halos may produce a neutrino signal from particle dark
matter annihilation, though this requires that the cusp continues inward
to 1 pc. Our experiments show that these neutrino signatures will not exist
for any galaxy that has gone through a barred phase, unless a supermassive
black hole has subsequently induced a cusp (Ullio et al 2002). In addition, spiral
structure can also couple to the dark-matter halo with the same mechanism.
Although spiral arms are weaker than bars, it is likely that modest
spiral structure in the inner disk would be capable of affecting cusps
at the parsec scale.

Bars are ubiquitously produced in galaxy simulations either through
local instabilities or tidal interactions (Barnes \& Hernquist 1992,
Noguchi 1996, Steinmetz \& Navarro 2002). Recent cosmological
simulations designed to track the morphological evolution of galaxies
predict that the bar phase is a natural byproduct of galaxy evolution
(Steinmetz \& Navarro 2002).  An $L_{\star}$ Sb galaxy at $z=0$ should
have experienced a large bar by $z=1.5$, and this should be observable
with NGST.  Both the higher galaxy gas fraction (Somerville, Primack,
\& Faber 2001) and the higher interaction rate (Le F\'evre et al 2000,
Kolatt et. al. 2000) should enhance high redshift bar
formation. Despite all this theoretical prejudice, the high redshift
bar fraction of SB galaxies is claimed to be smaller than the local
bar fraction: $5 \%$ for $0.6 < z < 0.8$ versus $30 \%$ in the local
Universe (van den Bergh et al. 2002, Abraham et al. 1999).  This might
be a selection effect.  For example, the identification of high
redshift bars might be hampered by low sensitivity (van den Bergh et
al.  2002).  Moreover, the classification technique used to identify
high redshift bars may misclassify bars undergoing strong starbursts
(Jogee et al. 2002), a characteristic event for a newly-formed bar in
a gas rich environment (Friedli \& Benz 1993, Sheth et al. 2002).  In
fact, recent NICMOS data have revealed that there is no significant
evidence for a decrease in the fraction of barred spirals out to 
$z \sim 0.7$ (Sheth et al 2003). Answers to this mystery will help define the
epoch of bar formation, constrain the ``duty cycle'' of bars and
specify the role of bars in driving galaxy evolution through the
mechanisms proposed in this paper.

\section{Summary}

Based on linear perturbation theory and idealized N-body simulations
for the evolution of a rotating bar in a cuspy dark-matter halo,
Weinberg \& Katz (2002) predicted that the bar slows as it loses
angular momentum to the halo and that the dark matter cusp flattens as it
gains angular momentum.  Resonant interactions
between halo orbits and the orbiting bar perturbation causes this
angular momentum transfer.  We have
performed high-resolution, self-consistent N-body simulations of
realistic stellar disks embedded in NFW dark matter halos and have
verified each of these predictions.  Our overall specific conclusions
are as follows:
\begin{enumerate}
  
\item The bar mediates significant angular momentum transfer in the
  galaxy, driving evolution in the disk and halo.  Approximately $30 \%$
  of the initial bar angular momentum is lost to the halo and outer
  disk after $t=2.7$ (3.5 Gyr, 6 Gyr).  The inner halo cusp is flattened to
  $R=0.003$ (170 pc for a dwarf, 900 pc for the Milky Way) and the inner
  disk scale length shrinks by $60 \%$ for our $R=0.05$ (870 pc, 4.5 kpc) bar.
  During this time, the disk transfers $16 \%$ of its initial total angular
  momentum to the halo through resonant interactions.
  
\item We empirically demonstrate the need for at least 5 million equal
  mass particles within the virial radius to correctly represent
  resonant dynamics using an expansion (SCF) N-body technique.  There
  are two reasons for this. First, with too few particles, Poisson
  fluctuations in the gravitational potential can overwhelm the
  resonance potential.  This is particularly important for the
  innermost ILR-like resonance that drives the cusp evolution.  As the
  particle number decreases, the resolved portion of this resonant
  potential lies further out in the halo, and the cusp evolution is
  suppressed.  Second, the resonant torque is an ensemble effect, and
  therefore requires many particles near the resonance in phase space.
  Paper II presents explicit numerical criteria for both conditions
  that agree with our empirical findings.  These particle number
  criteria are likely to be lower limits for potential solvers such as
  direct-summation, grid, or tree codes.  These criteria also ignore
  the effects of orbit diffusion, which would further increase the
  required particle number for these methods.
  
\item A similarly shaped bar forms after stimulation by an external
  trigger or through an internal instability.  The gravitational
  potential of either bar is well-represented by that of a homogeneous
  ellipsoid.  The shape and amplitude of the bar's quadrupole
  determines the radial scale over which resonant coupling can
  occur. A bar with a more gradual profile, such as that used by
  Hernquist \& Weinberg (1992) and by Sellwood (2003), provides a
  quadrupole thirty times smaller than that found here and than that
  used in Weinberg \& Katz (2002), explaining the negligible evolution
  and slow down reported by Sellwood (see Paper II for details).

\item We agree with some aspects of previous simulations of bar
  formation and evolution. We attribute the differences to either the
  use of unrealistic initial conditions or to small-scale numerical
  noise that artificially removes the dynamically important
  resonances.
  
\item We have shown that a tidal interaction can induce strong bars to
  form, grow, and persist at many disk scale lengths, well beyond the rising
  part of the galaxy's rotation curve.  Hence, the formation of
  large bars are compatible with flat rotation curve galaxies.

\item Bars, and other more general excitations, may reshape the halo
  structure for both high- and low-surface brightness galaxies and
  may affect a galaxy's morphological history.  This complicates
  the interpretation of present-day rotation curves as probes of the
  primordial dark matter halo distribution.

\end{enumerate}
\acknowledgements

KHB would like to thank Chris Mihos and Steinn Sigurdsson for
helpful comments on an earlier version of this draft.
This work was supported in part by NSF AST-0205969 and AST-9988146, and  
by NASA ATP NAG5-12038 and LTSA NAG5-13102.

\end{document}